\documentclass[a4paper,12pt]{JHEP3}
\pdfoutput=1

\usepackage{latexsym}
\usepackage{amsmath}
\usepackage{amsfonts}
\usepackage{mathrsfs}
\usepackage{amssymb}
\usepackage{dsfont}
\usepackage{slashed}
\usepackage{stmaryrd}
\usepackage{graphicx}

\newcommand{\La}{{\mathcal{L}}}
\newcommand{\ca}[1]{{\mathcal{#1}}}

\newcommand{\unit}{\mathds{1}}  
\newcommand{\fr}[2]{{\textstyle{\frac{#1}{#2}}}}
\newcommand{\D}[1]{\dot{#1}}
\newcommand{\del}{\partial}
\newcommand{\vep}{\varepsilon}
\newcommand{\vphi}{\varphi}
\newcommand{\vth}{\vartheta}

\newcommand{\ww}{\underline}

\newcommand{\Nf}{N_f}
\newcommand{\sN}{{\mathcal{N}}}
\newcommand{\aux}[1]{{\mathsf{#1}}}
\newcommand{\sdag}{\dagger}

\author
    {%
    David Burke$^1$ and Robert Wimmer$^2$    \\

   $^1$Institut f\"ur Theoretische Physik, Technische Universit\"at Wien,\\ Wiedner Hauptstr. 8-10,
       A-1040 Vienna, Austria  \\

$^2$Universit\'e de Lyon, Laboratoire de Physique, UMR 5672, CNRS, \\
\'Ecole Normale Sup\'erieure de Lyon,\\
46, all\'ee d'Italie, F-69364 Lyon cedex 07, France \\

{\tt dburke@hep.itp.tuwien.ac.at}, \  
{\tt robert.wimmer@ens-lyon.fr}
}

\title{Quantum Energies and Tensorial Central Charges of Confined Monopoles}

\dedicated{Dedicated to the memory of Max Kreuzer} 

\abstract{We study different aspects of monopoles in the Higgs phase which are confined by 
(non-abelian) vortices in $\sN=2$ SQCD with gauge group $U(N)$ and $N_f\geq N$ massive flavors, including 
generalized FI-terms.
We compute in particular the perturbative quantum corrections for (multiple)
confined monopoles and identify an anomalous contribution in the central charge.
For $N_f=N$ the results match the quantum corrections for kinks in two-dimensional $CP^{N-1}$ models. 
To regulate the theory we embedded it in a six-dimensional model with constant $U(1)$ background fields
which generate the masses upon dimensional reduction. We discuss the (local) susy algebra
and its representations including tensorial central charges, which are carried by the confined monopoles,
in an $SU(2)_R$ covariant way. The resulting $SU(2)_R$ covariant 1/4 BPS equations show that there
is no correlation between the $SU(2)_R$  and the spatial orientation of the confined monopole system.
However, at the quantum level we find that supersymmetry links the spatial $SU(2)$ with the
R-symmetry $SU(2)_R$.}      


\preprint{TUW-11-18}

\begin{document}

\section{Introduction}

As is understood by now, magnetic monopoles form a crucial part of the spectrum 
of many gauge theories and are essential for the understanding of non-perturbative properties. 
Though no magnetic monopole has been observed so far, their theoretical relevance
qualifies them as ``one of the safest bets'' \cite{Polchinski:2003bq} for yet unseen physics.

A phenomenon of utmost importance for the understanding of our world,  
where magnetic monopoles are expected to play a key role, is the 
confinement of quarks. A mechanism which is responsible for 
permanent quark confinement was proposed by 't Hooft \cite{'tHooft:1975pu}
and Mandelstam \cite{Mandelstam:1974vf}: The condensation of magnetic monopoles leads to
the formation of chromo-\emph{electric} flux tubes which connect and confine the
electrically charged quarks. This is a ``dual Meissner effect'', i.e. the electric-magnetic
dual scenario to the formation of chromo-\emph{magnetic} flux tubes which form 
when electrically charged (Higgs) fields/particles condense  and confine
magnetic monopoles. 
See \cite{'tHooft:1998wf} for a review of the development of these ideas. 
This picture gave a qualitative 
meaning to the mechanism behind confinement in QCD but also indicated that (analytic) quantitative
predictions are out of reach. However, this qualitative understanding implies that
electric-magnetic duality may play an important part in such considerations.

The situation is more promising in supersymmetric theories which became especially clear 
through the seminal work of Seiberg and Witten \cite{Seiberg:1994rs, Seiberg:1994aj}.
The explicit solution for the strongly coupled low-energy theory of (softly broken)
$\sN=2$ gauge theories allowed an analytic and exact description of the 't Hooft - Mandelstam 
mechanism of quark confinement, which is 
triggered by the condensation of monopoles and the formation of vortices. 
However, it was quickly realized that confinement as described in 
\cite{Seiberg:1994rs, Seiberg:1994aj} has phenomenologically unacceptable 
properties, even when considered as a toy model. The numerous vortices, one for
each $U(1)$ factor of the broken group, produce a hadron spectrum which is far too rich
\cite{Douglas:1995nw, Hanany:1997hr}. It is believed that the reason for this is that the solutions of 
\cite{Seiberg:1994rs, Seiberg:1994aj} describe abelian confinement where the 
gauge group is completely (dynamically) abelianized at strong coupling.

In \cite{Argyres:1996eh, Carlino:2000uk} it was observed that there exist certain 
vacua in $\sN=2$ theories,
so called ``$r$-vacua'', 
which survive the soft breaking to $\sN=1$ supersymmetry and preserve a non-abelian subgroup
$SU(r)$ at strong coupling, i.e. dynamical abelianization does not take place. 
Based on observations made in \cite{Marshakov:2002pc} it was then shown in 
\cite{Auzzi:2003fs} that for the associated symmetry breaking pattern
the theory admits truly non-abelian vortices.\footnote{This means that the vortices are not
merely abelian vortices embedded in a fixed direction of the Cartan subalgebra.} Around the same 
time a novel set of BPS-equations were found \cite{Tong:2003pz} that besides non-abelian vortices 
describe monopoles in the Higgs phase which are confined by these vortices. 
These  findings triggered tremendous developments recently in the study of non-abelian confinement. 
Subsequently it was shown that the effective low energy theory on the world sheet of 
such a non-abelian vortex is a $CP^{n}$ sigma model with twisted mass terms 
\cite{Hanany:2004ea, Shifman:2004dr} and that the kink solutions of these models 
correspond to confined monopoles \cite{Shifman:2004dr}.  This
also explained the observed matching between the BPS spectra of $\sN=(2,2)$ $CP^{N-1}$ models
and four-dimensional $\sN=2$ supersymmetric $SU(N)$ gauge theories on the coulomb branch with $N_f=N$ 
\cite{Dorey:1998yh}. For some recent reviews 
of these topics see \cite{Tong:2008qd, Konishi:2007dn,Shifman:2009zz}
\\

\noindent
{\bf{Dynamical Setting:}} One starts with $SU(N_c)$ $\sN =2$ SQCD with $\Nf$ massive flavors
and $\Nf < 2 N_c$, so that the theory is asymptotically free.
The mentioned $r$-vacua \cite{Argyres:1996eh, Carlino:2000uk} are characterized 
by an expectation value of the adjoint scalar at energies well above the dynamically 
generated scale $\Lambda$, 
i.e. $\langle\phi\rangle \gg \Lambda$, which breaks the gauge group in the pattern 
$SU(N_c)\rightarrow SU(r)\times U(1)^{N_c-r}$. In addition, these vacua are also $\sN=1$ vacua, i.e. 
they survive the soft breaking $\sN=2$ to $\sN=1$ by an adjoint mass term 
$\mu\, \mathrm{Tr}\, \Phi^2$.
For $2r\leq \Nf$ the non-abelian gauge coupling is frozen at the small value 
$\frac{1}{g^2}\sim \log(\frac{\langle\phi\rangle}{\Lambda})$. Therefore the non-abelian gauge group
does not dynamically abelianize as one flows to lower energies. 
The observation made in \cite{Auzzi:2003fs} is that for
the given window  $2 N_c > \Nf\geq 2 r$ perturbative reasoning for the effective 
$SU(r)\times U(1)^{N_c-r}$ theory is justified and that the a priori abelian vortices, which exist
after a further breaking of the gauge group by an expectation value of the Higgs scalars of the 
quark multiplet, are in fact degenerate states of a non-abelian vortex. The crucial condition for 
this is that the Higgs expectation value preserves a diagonal subgroup of the gauge and flavor 
symmetries. This is the  so called color-flavor locked phase. See \cite{Auzzi:2003fs} and the 
following section for further details.

The findings of \cite{Auzzi:2003fs} provide a possibility for the description of non-abelian
confinement by a 't Hooft - Mandelstam mechanism. 
Further evidence in this direction is given by the observation that the dual quarks are identified  
with GNO monopoles of the microscopic theory \cite{Bolognesi:2002iy}. 
In this paper we study different aspects of monopoles in the Higgs phase, confined by the vortices,
in the setting just described. The focus is on 
supersymmetry properties and the role played by the $SU(2)_R$ symmetry in this context and
especially on quantum properties - in particular the quantum energies of the confined monopoles and the 
comparison with $CP^{n}$ models. We also give a brief discussion of the central charge anomaly. 
We consider $\sN =2$ SQCD with gauge group $U(N)$ and $\Nf\geq N$ massive flavors, which corresponds 
to the case described above with $r=N=N_c-1$, where for notational convenience we
take the gauge group to be $U(N)$ instead of $SU(N)\times U(1)$. The physical results are 
essentially the same. For small $\mu$ the soft-breaking term 
$\mu \, \mathrm{Tr}\, \Phi^2$ can be approximated by an $F$-term which preserves $\sN=2$ supersymmetry 
and is a member of a triplet of generalized  Fayet-Iliopoulos terms which we include. 
For a reliable perturbative analysis of the low energy scenario described above 
one would have to choose $\Nf \geq 2N$ in the end, but formally the calculations are independent of
this restriction. We will in general not make reference to the ambient 
scenario described here but will consider the theory with its own parameter space, independent of its possible 
origin. However, the ambient scenario that we just discussed guarantees that for certain 
choices of the parameters our results are valid also for the low energy dynamics.

\begin{figure}[tp]
 \label{fig}
  \centering
  \includegraphics[width=15cm]{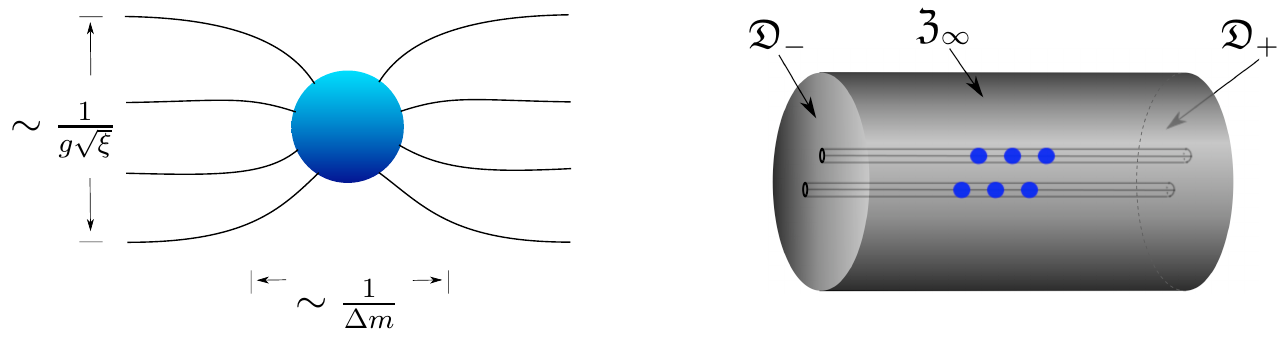}
  \caption{The left figure shows a confined monopole in close up with the characteristic
    scales for the monopole size displaying approximate Coulomb monopole behavior 
    and the size of the confining vortex. The right figure depicts the spatial boundary, a cylinder at infinity,
    where the asymptotic behavior of the (multiple) confined monopoles is specified (see main text). The cylinder 
    at infinity consists of the two \emph{discs} $\mathfrak{D}_{\pm}$ at $x^3\rightarrow\pm \infty$ and the 
    \emph{cylinder wall} $\mathfrak{Z}_{\infty}$ at $r\rightarrow\infty$, with $r$ being the  
    cylindrical radial coordinate.}
\end{figure}

The geometrical setting of the confined monopolies is assumed to be axially oriented 
in the direction of the $x^3$-axis as illustrated in figure \ref{fig}. Such configurations 
preserve $\frac{1}{4}$ of the original supersymmetry, as will be discussed below.

The paper is organized as follows: In section 2. we describe three different 
formulations of the model to discuss different properties. In particular
we introduce an embedding in a six-dimensional theory which is crucial for 
the quantum computations. In section 3. we discuss the super-multiplet structure 
of confined monopoles. To do so we identify tensorial central charges 
in the susy algebra. In addition we discuss the $SU(2)_R$ action on the associated BPS equations. 
In section 4. we perform the quantization of the confined monopoles and compute 
the quantum energies for these objects, identify an anomaly in the central charge
and compare our results with the BPS spectrum of $CP^{n}$ models. 
In section 5. we summarize our results and present
conclusions.

\section{Three Faces of ${\cal N}=2$ SQCD}

In this section we give three different formulations of ${\cal N}=2$ SQCD with 
$\Nf$ massive quark multiplets and a generalized Fayet-Iliopoulos FI-term.
Each of the three formulations will allow us to analyze different aspects in a useful way.
First we introduce a $\sN=1$ superspace formulation to list some basic properties of the 
model and in particular discuss the vacua and symmetry breaking patterns in a conventional 
way. Secondly we give an $SU(2)_R$ covariant component formulation which
will set the conventions for the subsequent sections and notably be the starting point 
for a six-dimensional formulation of the theory. Lastly we construct a six-dimensional 
model in which the four-dimensional theory can be embedded. This formulation is of
utmost importance for the consistently regularized quantization procedure in 
the presence of monopoles and vortices. 
The gauge group is  chosen to be $U(N)$ and in general we assume that $\Nf \ge N$,
as discussed in the introduction.

\subsection{${\cal N}=1$ Superspace and Vacuum Moduli}\label{secvac}

The ${\cal N}=2$ SQCD Lagrangian is given by the coupling of a ${\cal N}=2$ 
vector multiplet in the adjoint representation to $\Nf$ hypermultiplets 
in the fundamental representation of the gauge group. In ${\cal N}=1$ 
superspace the ${\cal N}=2$ vector multiplet is composed of a vector multiplet 
$V\sim (A_\mu,\lambda, D)$ and a chiral multiplet 
$\Phi\sim (\phi,\psi,F)$, both in the adjoint representation. The $i=1,\ldots,\Nf$  hypermultiplets 
are composed of two sets of chiral multiplets $Q_i \sim(q_i, \chi_i, F_i)$
and $\tilde{Q}_i \sim(\tilde{q}_i, \tilde{\chi}_i, \tilde{F}_i)$, where 
$Q_i$ transforms in the fundamental ${\bf N}$ of $U(N)$, whereas  $\tilde{Q}_i$
transforms in the conjugate representation  ${\bf \bar N}$. Similarly, w.r.t. 
the flavor group $SU(\Nf)$ the multiplets  $Q_i,  \tilde{Q}_i$ are in the
representation ${\bf N_f}$ and ${\bf \bar N_f}$, respectively, 
though we do not indicate this by the index position. We do not include a vacuum theta-angle.
The superspace Lagrangian is given by\footnote{We essentially use Wess and 
Bagger conventions \cite{Wess:1992cp} except that we contract 
2-component spinors with $\vep^{12}=\vep_{12}=1$ so that 
$\lambda^\alpha=\vep^{\alpha\beta}\lambda_{\beta}\ \leftrightarrow\
\lambda_\alpha=\lambda^{\beta}\vep_{\beta\alpha}$. The superfield strength 
is defined as $W_\alpha=-\frac{1}{4}\, {\bar D}^2\, e^{2V}D_\alpha e^{-2V}$. Generally summation 
over repeated flavor indices is implied. For the 
metric we use east coast convention $\eta_{\mu\nu}=(-,+,+,+)$.}
\begin{align}
  \label{eq:lss}
  \La =&\ \ \mathrm{Tr}\ \big\{\, \tfrac{2}{g^2}  
      \int d^{2}\theta\, W^{\alpha} W_{\alpha}+h.\,c. + 
       \tfrac{2}{g^2}\int d^{4} \theta\, e^{2V} \Phi^{\dagger} e^{-2V} \Phi 
      \nonumber \\
      &\hspace{4cm} +2\,\xi_3 \int d^{4} \theta\, V -
       \fr{i}{\sqrt{2}}\, \bar{\xi}  \int d^{2} \theta \, \Phi +h.\,c.\big\}
          \nonumber\\[4pt]
     & + \int d^{4} \theta\, (\, \bar{Q}_{i}\, e^{-2V} Q_{i} + 
          \tilde{Q}_{i}\, e^{2V} \bar{\tilde{Q}}_{i}\, ) +
      \sqrt{2} \int d^{2}\theta\,(\, \tilde{Q}_{i} \Phi\, Q_{i} + m_{i} 
        \tilde{Q}_{i} Q_{i} +h.\,c.).
\end{align}
The first line describes pure  ${\cal N}=2$ SYM 
and the second line resembles a generalized FI-term and is non-vanishing 
for $U(N)$ or any $U(1)$ factor in the considered gauge group. The first FI-term is a
standard FI-$D$-term whereas the second term, the non-standard FI-$F$-term,
 can be obtained from a 
soft breaking mass term for the adjoint chiral field $\Phi$ which to first order 
preserves ${\cal N}=2$ supersymmetry \cite{Hanany:1997hr, Vainshtein:2000hu}. 
See also the comments in 
the introduction. The FI-parameters $\xi_3$ and $\xi=\xi_1+i \xi_2$ form 
an $SU(2)_R$ vector (triplet)  $\vec{\xi}$ \cite{Shifman:2009zz} which explicitly breaks 
the $SU(2)_R$ symmetry. At the considered scale 
within the mentioned approximation this allows one to interpret
all FI-terms to be of dynamical origin due to a soft breaking term. 
Different choices
 for $\vec{\xi}$, including the standard FI-$D$-term which is usually put in by hand,
are related to each other by $SU(2)_R$ transformations. We will make use of this in more detail
below. This is similar to the explicit breaking of the flavor symmetry 
$SU(\Nf)$ by the masses $m_i$ which was used to bring the mass matrix into diagonal form (for $\sN =2$ 
supersymmetry the mass matrix has to commute with its hermitian conjugate 
\cite{Argyres:1996eh}).    

The last line is the Lagrangian for the 
quark multiplets, which turns pure SYM into SQCD. The superpotential contains 
a priori complex masses $m_i$ 
whose specific values are crucial for the resulting dynamics. Eventually we will choose them to be real and they can be ordered by $SU(\Nf)$ Weyl reflection such that $m_1\leq m_2 \leq \ldots \leq m_{\Nf}$.  
Their relative 
values determine the symmetry breaking pattern of the gauge group as well as the explicit breaking 
of the $SU(\Nf)$ flavor symmetry. For
generic masses $U(N)$, $SU(\Nf)$ is broken to $U(1)^{N}$, $U(1)^{\Nf-1}$ at the scale of the 
bare masses  $m_i$, which are defined at a UV cut-off scale $M_{UV}$.\footnote{The baryonic 
$U(1)$ and the explicitly broken $U(1)_R$ 
symmetry of the model will be of less relevance in the following.} 
The coefficient for the superpotential is chosen such that the theory 
has ${\cal N}=2$
supersymmetry, there is no independent coupling for it.

For the generators of the gauge group $U(N)$ we have the following conventions: 
The hermitian generators are $\{T^A\}=\{T^0,T^a\}$, where $\{T^a\}$ forms an $su(N)$ 
algebra, and  satisfy
\begin{equation}
  \label{eq:un}
  [T^a,T^b]=i f^{abc}\, T^c\ ,\ \ T^0=\fr{1}{\sqrt{2 N}}\unit\ , 
     \ \ \mathrm{Tr}\{T^A T^B\}=\fr{1}{2}\,\delta^{AB}\ ,
\end{equation}
where $f^{abc}$ are the real and totally antisymmetric $su(N)$ structure constants.
\\

\noindent
{\bf Classical Vacua.} The Lagrangian (\ref{eq:lss}), though  only
explicitly realizing $\sN=1$ supersymmetry,  is a convenient way 
to discuss the vacuum moduli. The presence of the FI-terms drives the 
the theory into the Higgs phase at the scale where they become relevant. The different 
possible vacua lead to different symmetry breaking patterns and associated topological 
field configurations. We discuss this in the following.

The bosonic potential of (\ref{eq:lss}) 
can be written in terms of the on-shell component auxiliary fields $D$, $F$ and $F_i$, $\tilde{F}_i$,
\begin{equation}
  \label{laux}
  - \La_{\mathrm{aux}} = V_{\mathrm{bos}}= \fr{2}{g^2}\,\mathrm{Tr}\,\{\fr{1}{2}\, D^2 + F F^\dagger\}
        +\sum_{i}\, |F_i|^2 + |\tilde{F}_i|^2 \, ,
\end{equation}
where the auxiliary fields are given by
\begin{align}
  \label{aux}
  & D = [\phi,\phi^\dagger ] 
    + \fr{g^2}{2} ( q_i\otimes\bar{q}_i - \bar{\tilde q}_i\otimes \tilde{q}_i -\xi_3 \unit) \ ,\  \
  F^\dagger = -\fr{g^2}{\sqrt{2}}\, ( q_i\otimes \tilde q_i -\fr{i}{2}\,  \bar{\xi}\,\unit) \nonumber \\
  &\hspace{1.5cm} F_i = -\sqrt{2} (\phi^\dagger + \bar m_i) \bar{ \tilde q}_i\ ,\ \
  \bar{\tilde F}_i = -\sqrt{2} (\phi + m_i) q_i\ ,
\end{align}
and there is no summation over flavor indices in the second line. Generally we employ a matrix 
notation for the adjoint fields, accordingly the tensor product in the first line is defined w.r.t. 
gauge indices. The vanishing of the potential (\ref{laux}) thus implies the conditions
\begin{align}
  \label{vac}
 & [\phi, \phi^\dagger] = 0\, , \nonumber\\ 
 &  q_i\otimes \tilde q_i = \fr{i}{2}\,  \bar\xi\,\unit \ \ , \ \ 
   q_i\otimes\bar{q}_i - \bar{\tilde q}_i\otimes \tilde{q}_i =  \xi_3 \unit\, ,\nonumber\\ 
 &(\phi + m_i) q_i = 0\ , \ \  (\phi^\dagger + \bar m_i) \bar{ \tilde q}_i = 0 \ , 
\end{align}
which are of the same form as for $SU(N)$ theory \cite{Argyres:1996eh} except for the 
absence of the condition
$\mathrm{Tr}\,\phi=0$ and that the FI-parameters $\vec{\xi}$ fix the moduli of 
the Higgs branch equations.
We refer to \cite{Argyres:1996eh} for further details, but here we are interested in a 
certain set of vacua
for which the squark fields $q_i$, $\tilde q_i$, written as 
$N\times \Nf$ matrices, take the form
\begin{equation}
  \label{vac1}
  [q^{\mathrm{vac}}]^{n}{}_i =  \begin{bmatrix}
                \kappa_1 &         &          &0&        &      & 0 &\\
                         & \ddots  &          &         & \ddots &      &   &\ddots  \\ 
                         &         & \kappa_N &         &        &0 & &
             \end{bmatrix} \ , \ \  
  [\tilde{q}^{\mathrm{vac}}]_{n i} = \begin{bmatrix} 
           \tilde\kappa_1 &         &          &0       &        &       & 0 &  \\
                         & \ddots  &          &         & \ddots &      &   &\ddots  \\ 
                         &         & \tilde\kappa_N &         &        & 0    &   &
             \end{bmatrix}\ ,
\end{equation}
where we assumed for the moment that $\Nf \ge 2 N$ and indicated that these are the vacuum values of the fields.
In general we will assume only $\Nf \ge N$, which 
is taken into account by erasing the appropriate number of columns in (\ref{vac1}). 
The Higgs branch equations, the second line in (\ref{vac}), give
\begin{equation}
  \label{vac2}
  |\kappa_n|^2-  |\tilde\kappa_n|^2 = \xi_3 \  \ \ \textrm{and} \  \ \ 
   \kappa_n \tilde\kappa_n =\fr{i}{2}\, \bar\xi\ , \ \ \forall\,  n =1,\ldots, N\ . 
\end{equation}

For certain solutions of (\ref{vac2}) the vacuum preserves the symmetry
$SU(N)_{C+F} = \mathrm{diag}(U(N)_C\times SU(N)_F)$. This 
symmetry of the so called color-flavor locked phase is the key 
aspect for the existence of non-abelian vortices \cite{Auzzi:2003fs}. We 
therefore set $\kappa_n =\kappa$
and $\tilde\kappa_n =\tilde\kappa$ for all $n$. 
Inserting in (\ref{vac1}) and deleting the trivial columns 
shows that 
\begin{equation}
  \label{cf}
    U_C \,q^{\mathrm{vac}} \,U_F = q^{\mathrm{vac}}\ \ ,\ \  
     U^\dagger_F\, \tilde{q}^{\mathrm{vac}}\, U^\dagger_C = \tilde{q}^{\mathrm{vac}}\ ,
\end{equation}
where $U_C\in U(N)$, $U_F \in SU(\Nf)$ are $SU(N)$ transformations with $U_C = U_F^{-1}=:U_{C+F}$. 

The actual values of $\kappa$, $\tilde\kappa$ depend on the FI-parameters and are determined by (\ref{vac2}).
There are two special choices which appear in the literature. i.)  $\xi =0$, $\xi_3 >0$ 
gives $\tilde{q_i}^{\mathrm{vac}}=0$, and $q^{\mathrm{vac}\, n}_i=\sqrt{\xi_3}\,\delta^n_i$. 
ii.) $\xi_3=0$, $\xi_1=0$, $\xi_2 >0$
gives $\tilde q^{\mathrm{vac}}_{n i}= q^{\mathrm{vac}\, n}_i = \sqrt{\xi_2/2}\,\delta^n_i$. 
The phases of $q_i^{\mathrm{vac}}$, 
$\tilde {q_i}^{\mathrm{vac}}$ have been fixed by gauge symmetry. A
nontrivial winding in these phases describes topologically stabilized configurations.
The classification of such states is given by the symmetry breaking pattern of the vacuum,
\begin{equation}
  \label{vac41}
  U(N)\times SU(\Nf) \overset{\sqrt{\xi}}{\longrightarrow} SU_{C+F}(N) \, ,
\end{equation}
where the breaking of the overall $U(1)\subset U(N)$ guarantees vortex solutions which are topologically 
stabilized according to $\pi_1(U(1))=\mathbb{Z}$ and are non-abelian for $N>1$ due to the 
preserved color-flavor symmetry $SU_{C+F}(N)$.

This is of course not the whole story since a vev for the adjoint scalar
may break the gauge group to a subgroup of $U(N)$ already at a higher scale, as we will 
assume in the following. 
For non-vanishing FI-parameters either $\kappa$ or $\tilde\kappa$ is non-vanishing and 
thus the last two conditions in (\ref{vac}) imply for the vev $\phi_0$ of $\phi$,
\begin{equation}
  \label{vac3}
       \phi_0 = -  \begin{bmatrix} 
                        m_1 &    & \\
                            & \ddots & \\
                            &       & m_N
                  \end{bmatrix}   \ ,  
   \quad \quad \textrm{with} \quad H_C\, \phi_0\, H^{-1}_C = \phi_0\ , 
\end{equation}
which thus automatically satisfies the first condition in (\ref{vac}). For generic masses
the gauge symmetry is completely broken ($SU(\Nf)$ is explicitly broken). However, if some 
masses coincide the unbroken subgroup $H_C \subset U(N)$ is non-abelian and also the 
flavor group $SU(\Nf)$ is not completely broken 
and hence part of the 
color-flavor symmetry (\ref{cf}) remains intact. Consequently, the symmetry breaking pattern is given by
\begin{equation}
  \label{vac4}
  U(N)\times SU(\Nf)  \overset{m}{\longrightarrow} U_C(1)\times H_C \times H_F 
  \overset{\sqrt{\xi}}{\longrightarrow} H_{C+F} \, ,
\end{equation}
where we assumed that all masses are of the same scale and that $m \gg \sqrt{\xi}$. If the first $N$ masses form $q$ groups of $n_r$ degenerate masses the surviving symmetry group is given by \cite{Nitta:2010nd}
\begin{equation}
  \label{HCF}
  H_{C+F}= S(\, U(n_1)\times \ldots \times U(n_q)\,)\, ,
\end{equation}
with $\sum_r n_r =N$. It supports monopoles with typical size $1/\Delta m$, the 
inverse mass difference, and
are confined by flux tubes of width $\sim 1/g \sqrt{\xi}$, see figure \ref{fig}.

\subsection{$SU(2)_R$ Covariant Formulation} 

The component form of the Lagrangian (\ref{eq:lss}) can be easily obtained by using the 
formulas given in \cite{Wess:1992cp}. Here we give the resulting component Lagrangian directly in the 
$SU(2)_R$ covariant form. For this one groups the adjoint fermions of the 
$\sN = 2$ vector multiplet $(V,\Phi)$  and the fundamental scalars of the quark multiplets 
$(Q_i, \bar{ \tilde Q}_i)$ into $SU(2)_R$ doublets $\lambda_{\alpha I}$ and $S_{i I}$, 
respectively:
\begin{equation}
  \label{sur1}
    \lambda_{\alpha I} := (\lambda_\alpha,\psi_\alpha) \ \ , \ \ 
    S_{i I} := (q_i, i \bar{\tilde q}_i)\ ,
\end{equation}
with $(\lambda_{\alpha I})^\dagger=: \bar\lambda^I_{\dot\alpha}$ and $(S_{iI})^*=:\bar S^I_i$.
With this notation the resulting component Lagrangian obtained from (\ref{eq:lss}) can be written
in the following form:
\begin{align}
  \label{lsur}
 \hspace{-0.5cm} \La_{4D}&=\nonumber\\
   &\fr{2}{g^2}\,\mathrm{Tr}\,\{ -\fr{1}{4} F_{\mu\nu}^2 - |D_\mu \phi|^2 
      - i\bar\lambda^I \bar{\slashed{D}}\lambda_I \}
      -  |D_\mu S_{i I}|^2
 - i (\bar\chi_i \bar{\slashed{D}}\chi_i + \bar{\tilde\chi}_i \bar{\slashed{D}} \tilde\chi_i)
   \nonumber \\[10pt]
 &-{\textstyle{\sqrt{2}}}\,\big{[}\, \fr{i}{g^2}\,\mathrm{Tr}\,\{\vep_{IJ}\, \bar\lambda^I 
                       [\, \bar\lambda^J, \phi]\} 
   +i \bar S^I_i (\lambda_I\chi_i - i\,\vep_{IJ}\bar\lambda^J \bar{\tilde\chi}_i )
    +\tilde\chi_i (\phi + m_i) \chi_i +h. c.\big{]} \nonumber\\[10pt]
 &-\fr{2}{g^2}\,\mathrm{Tr}\,\{\,
       \fr{1}{2}[\phi,\phi^{\dagger}]{}^2 -\fr{1}{2}\,{\vec{\aux{D}}}^2 + 
     \fr{1}{2}\,g^2\, \vec{\aux{D}}\cdot(\vec{\tau}_I{}^J S_{i\,J}\otimes \bar S^I_i
     - \vec{\xi}\ \unit )\} 
      \nonumber \\
     &\hspace{7.5cm}-\bar S^I_i\{\phi^\dagger+\bar m_i,\phi+m_i\}S_{i\,I}\ ,
\end{align}
where the first line contains the kinetic terms with\footnote{The covariant derivative is defined 
as $D_\mu=\partial_\mu-i A^R_\mu$ for fields in the representation $R$. The field strength is
given by $F_{\mu\nu}=\del_\mu A_\nu-\del_\nu A_\mu-i[A_\mu,A_\nu]$.}  
$ \bar{\slashed{D}} = \bar\sigma^\mu D_\mu$, 
and the second line gives the Yukawa couplings. The $SU(2)_R$ $\vep$-tensor is defined as 
$\vep^{12} = \vep_{12}=1$ and we use the convention $(\vep_{IJ})^*=(\vep^{IJ})$.\footnote{We do not raise/lower
$SU(2)_R$ indices with $\vep^{IJ}$, $\vep_{IJ}$, but index positions are changed by complex conjugation.} 
The residual terms
are the bosonic potential terms 
where the braces in the last term denotes the anti-commutator of the given matrices.

Here we introduced an $SU(2)_R$ triplet of auxiliary fields $\vec{\aux{D}}$ 
and the $SU(2)_R$ Pauli matrices $\vec\tau$. The 
introduction of these 
auxiliary fields does not result in an off-shell closure of the supersymmetry algebra
\cite{Sohnius:1985qm} but considerably simplifies many manipulations.
On-shell the auxiliary triplet is given by
\begin{equation}
  \label{d}
  \vec{\aux D}= \frac{g^2}{2}\, (\vec{\tau}_I{}^J S_{i\,J}\otimes \bar S^I_i - \vec{\xi}\ \unit )\ .
\end{equation}

This form of the Lagrangian is, up to the FI-vector $\vec{\xi}$, manifestly invariant under 
the $SU(2)_R$ symmetry which acts
on the adjoint fermion  doublet $\lambda_{\alpha I}$ and the squark scalar doublet $S_{i I}$ 
(and of course the auxiliary triplet $\vec{\aux D}$). All other fields are singlets under 
$SU(2)_R$. 

It has turned out that the embedding of a theory with solitons into
higher dimensions is a
crucial step in computing quantum corrections for (topologically) nontrivial states 
in a tractable and consistent way \cite{Goldhaber:2004kn, Rebhan:2004vn}. 
In the next step we will embed $\sN=2$ SQCD
with masses and FI-terms in six-dimensional space. This embedding and the relation
to four-dimensional physics is most conveniently found starting from the four-dimensional 
Lagrangian in the form (\ref{lsur}).

\subsection{Six-Dimensional Embedding} 

As is well known, pure $\sN=2$ SYM in four dimensions can be obtained by dimensional reduction from 
pure $\sN=1$ SYM in six dimensions \cite{Brink:1976bc}. In extending this method to  $\sN=2$ SQCD
one has to handle the masses and the extra field content from the quark hypermultiplet.

We first introduce the conventions for six-dimensional fermions, where we follow 
closely \cite{Rebhan:2006fg}.
We label coordinates by $x^M$, $M=0,1,2,3,5,6$, where $x^5$, $x^6$ are the extra dimensions
compared to the four-dimensional world. The six-dimensional gamma matrices $\Gamma^M$ 
and the chirality and charge conjugation matrices $\Gamma_7$, $C$ we define as\footnote{The metric
is $\eta_{MN}=(-1,1,\ldots,1)$ and in six dimensions one has $C^T=C$, and $C^\sdag = C^{-1}$.}  
\begin{align}
  \label{6dG}
  \Gamma_7 := \Gamma_0\Gamma_1\Gamma_2\Gamma_3\Gamma_5\Gamma_6 \ , \ \ 
   \Gamma_7{}^2= \unit \  , \ \  \Gamma_7{}^\sdag = \Gamma_7 \ , \ \ 
  C\, \Gamma^M C^{-1} = -{\Gamma^M}^T.
\end{align}
For the fermions we introduce symplectic Majorana-Weyl spinors $\lambda_{I=1,2}$, which 
are in the adjoint 
of the gauge group, and $\Nf$ anti-chiral spinors $\psi_{i=1\ldots\Nf}$ which are in the 
fundamental representation. These are eight-component spinors in six dimensions and satisfy
\begin{equation}
  \label{6dspin}
  \Gamma_7\,\psi_i = -\psi_i\ \ , \ \ \Gamma_7\,\lambda_I=\lambda_I\ \ ,\ \   
    \lambda_I=\vep_{IJ}\,C^{-1}\,{\bar{\lambda}}^{J^T} \ .
\end{equation}
Here the bar stands for Dirac conjugation, ${\bar{\lambda}}^J=(\lambda_J)^\sdag \,i \Gamma^0$, 
and the symplectic Majorana-Weyl condition can be equivalently written as
 ${\bar{\lambda}}^I=-\vep^{IJ}\lambda_J^T C$.

Next we introduce a $U(N)$ gauge field $A_M$ in six dimensions whose extra components give 
the adjoint scalars upon dimensional reduction. To also generate the masses for the (anti) fundamental 
matter fields in (\ref{lsur}) we introduce additional constant $U(1)$ background gauge fields 
in the extra dimensions, $\hat A_5$ and $\hat A_6$, so that $A_M=(A_\mu,\,A_5+\hat A_5,\,A_6+\hat A_6)$. 
The $i$'th flavor couples to these $U(1)$ 
backgrounds with charges 
$e_{5\,i}$, $e_{6\,i}$ if it is in the fundamental representation and with the negative thereof 
if it is in the anti-fundamental representation of the gauge group. Hence one has for example
\begin{equation}
  \label{dm}
  D_{5/6}\psi_{i}=\partial_{5/6}\psi_i-i(A_{5/6}+ e_{5\,i/6\,i})\psi_i\ .
\end{equation}
We introduce the same set of scalars $S_{iI}$ and their conjugates in six dimensions as in 
(\ref{lsur}), which also couple to the constant $U(1)$ backgrounds $\hat A_5$, $\hat A_6$ in the described 
manner, i.e. analogous to (\ref{dm}). As in four dimensions we define the covariant derivatives
of the adjoint fields and the six-dimensional field strength as
\begin{align}
  \label{f6d}
  F_{MN} &= \partial_M A_N - \partial_N A_M -i [A_M, A_N] \ , \nonumber\\
  D_{M}\lambda_I &= \partial_{M}\lambda_I - i [A_M, \lambda_I] \ .
\end{align}
Clearly, the constant $U(1)$ backgrounds enter exclusively through the 
coupling (\ref{dm}). 

The six-dimensional $\sN=1$ SQCD Lagrangian with FI-term is
\begin{align}
  \label{l6d}
  \La_{6D}=\fr{2}{g^2}\,\mathrm{Tr}\,\{& -\fr{1}{4} F_{MN}^{\, 2} - 
         \fr{1}{2} \bar{\lambda}^{I} \slashed{D} \lambda_{I} +\fr{1}{2}\,{\vec{\aux{D}}}^2 -
     \fr{1}{2}\,g^2\, \vec{\aux{D}}\cdot(\vec{\tau}_I{}^J S_{i\,J}\otimes \bar S^I_i
     - \vec{\xi}\ \unit )\} 
     \nonumber\\[5pt]
    & -|D_MS_{iI}|^2 - \bar\psi_i\slashed{D}\,\psi_i 
      +i\sqrt{2}\,(\, \vep_{IJ}\,\bar{S}^I_i\,\bar{\lambda}^J\psi_i 
     + \vep^{IJ}\,\bar\psi_i\,\lambda_I S_{iJ})\ ,
\end{align}
where $\slashed{D} =\Gamma^MD_M$, keeping in mind the additional $U(1)$ backgrounds. 
The six-dimensional Lagrangian (\ref{l6d}) is constructed in 
such a way that upon trivial dimensional reduction, i.e. by taking all fields independent of 
$x^5$ and $x^6$, one obtains the four-dimensional Lagrangian (\ref{lsur}).
\\

\noindent
{\bf Dimensional Reduction.} The first step is to choose a representation for the 
six-dimensional gamma matrices $\Gamma^M$ in terms of four-dimensional gamma matrices
$\gamma^\mu$ and decompose the spinors accordingly. Following \cite{Rebhan:2006fg} we choose
\begin{align}
  \label{6d4d}
  \Gamma^{\mu} &= \gamma^{\mu} \otimes \sigma^{1}\ ,\ \ \Gamma_{5} = \gamma_{5} \otimes \sigma^{1}
   \ , \ \ \Gamma_{6} = \unit \otimes \sigma^{2} \ , \nonumber\\
  C &=  C_{4D} \,\gamma_{5} \otimes \sigma^{2}\ , \ \ \Gamma_{7} = \unit \otimes \sigma^{3}\ ,  
\end{align}
where the form of $\Gamma_7$ follows from its definition in (\ref{6dG}) which also implies 
that\footnote{In four dimensions one has $C_{4D}^T=-C_{4D}$.} 
$\gamma_5= i \gamma_0\gamma_1\gamma_2\gamma_3$. The chirality conditions (\ref{6dspin})
thus imply that the six-dimensional eight component spinors are decomposed into  
four-dimensional four-component spinors as
\begin{equation}
  \label{6d4dspin}
  \lambda^{6D}_I = \lambda^{4D}_I\otimes \textstyle{\binom{1}{0}} \ \ ,\ \ 
   \psi^{\,6D}_i = \psi_i^{\,4D}\otimes \textstyle{\binom{0}{1}}\ , 
\end{equation}
where in the following we will omit the indication of the dimension; it will be clear from the
context.\footnote{The six-dimensional symplectic Majorana-Weyl condition in (\ref{6dspin}) translates
into $\lambda_I=-i\,\vep_{IJ}\gamma_5 C^{-1}\,\bar{\lambda}^{J^T}$ for the four-dimensional spinors. 
This is a symplectic Majorana condition \cite{Sohnius:1985qm}. We will not make
use of this description, most naturally the $SU(2)_R$ symmetry is formulated with two-component 
Weyl spinors as in (\ref{lsur}).} To make contact with our four-dimensional $\sN=2$ SQCD 
Lagrangian (\ref{lsur}) we choose a chiral representation for the four-dimensional gamma matrices 
$\gamma^\mu$ and decompose the four-component spinors into Weyl spinors as follows:
\begin{equation}
  \label{4dchiral}
  \gamma^\mu = i \begin{pmatrix} 0 & \sigma^\mu\\ \bar\sigma^\mu&0\end{pmatrix}\ \ , \ \ 
 \lambda_I = \begin{bmatrix} -i\,\lambda_{\alpha I} \\ \vep_{IJ}\bar\lambda^{\dot \alpha\, J}\end{bmatrix} 
  \ \ , \ \ 
    \psi_i =  \begin{bmatrix}\, \chi_{\alpha i}\, \\ \,\bar{\tilde\chi}_i^{\dot \alpha}\,\end{bmatrix} \ ,
\end{equation}
and the charge conjugation matrix is given by $C_{4D}=i\,\gamma^2\gamma^0$.
Inserting these decompositions into the six-dimensional Lagrangian (\ref{l6d}) and taking
all fields independent of the extra coordinates $x^{5,6}$ one finds
the original four-dimensional Lagrangian (\ref{lsur}) upon the following identifications:
\begin{equation}
  \label{A56}
  \phi = \frac{1}{\sqrt{2}}\, (i\, A_5 - A_6)\ \ ,\ \ m_i =  \frac{1}{\sqrt{2}}\, (i\, e_{5\,i} - e_{6\,i})
   \  .
\end{equation}
The mechanism introduced here for generating the masses in the lower dimensional theory 
is rather different from the usual Scherk-Schwarz mechanism \cite{Scherk:1979zr}. It is similar to
the method used in gauged $CP^{n}$ sigma models to generate twisted mass terms via 
constant background gauge fields \cite{Hanany:1997vm}. Nevertheless it would be interesting to have a brane
interpretation of this mechanism. 

To obtain the four-dimensional Lagrangian (\ref{lsur}) we consequently had to choose a chiral 
representation for the four-dimensional gamma matrices. However, for determining quantum corrections
for confined monopoles we will have to use a different representation to make use of the 
flat extra dimensions as a regulator. This will be discussed in more detail below.

\section{Tensorial Central Charges and $\frac{1}{4}$ BPS Equations}

In this section we analyze the supersymmetry structure of $\sN=2$ SQCD, especially the 
influence of the FI-terms. In particular we will find tensorial central charges in 
the local superalgebra (or in the global algebra in compact space). Usually 
such charges are not considered in the $\sN=2$ susy algebra, though they are common for
example in the M-theory superalgebra \cite{Townsend:1997wg}, but they are essential for the objects 
under investigation (see however recent developments in 
the context of six-dimensional $(2,0)$ theories compactified to four dimensions \cite{Moore}
and considerations for $\sN=1$ theories \cite{Dvali:1996xe,Gorsky:1999hk}).  
These tensorial central charges are carried by extended objects
such as the vortices which confine the monopoles or domain walls. The relation of 
these structures to the $\frac{1}{4}$ BPS equations, which also describe confined monopoles, will be 
described in the following.

\subsection{Supercurrent and SUSY Algebra}

It will be most convenient to start in the six-dimensional setting (\ref{l6d}). The Lagrangian
(\ref{l6d}) is invariant up to total derivative terms under the following susy transformations
\begin{align}
  \label{dsusy}
  \delta \vec{\aux{D}} &= - i\, \bar{\eta}^I \vec{\tau}_I{}^J\, \slashed{D}\, \lambda_J
    \hspace{3.5cm}\delta S_{iI}  =  - \sqrt{2}\, \vep_{IJ}\, \bar{\eta}^J\, \psi_{i}  \nonumber\\ 
  \delta A_M &= \,\bar{\eta}^I \Gamma_M \lambda_I 
   \hspace{4.6cm} \delta \psi_{i} = - \sqrt{2}\, \vep^{IJ} \slashed{D} S_{Ii} \eta_{J} 
     \nonumber \\ 
  \delta \lambda_I& = -\fr{1}{2}\, F_{MN} \Gamma^{MN}\, \eta_I 
     + i\, \vec{\aux{D}}\cdot \vec{\tau}_I{}^J \eta_J \ ,
\end{align}
where in the left column are the standard transformations of the gauge multiplet supplemented 
by the transformation of the adjoint auxiliary triplet and on the right are the 
transformations of the matter multiplet. Obviously the six-dimensional language and 
the use of the auxiliary field crucially simplifies the structure. The transformation parameter
$\eta_I$ is (like $\lambda_I$) a chiral symplectic Majorana-Weyl spinor and satisfies the relations 
given in (\ref{6dspin}). 

To determine the associated supercurrent one can use different versions of the Noether theorem, 
for example transforming the Lagrangian with local susy parameters. However, every procedure 
has some ambiguity in the resulting current. See for example \cite{Rebhan:2006fg} and the relevance of
improvement terms in this context. We determine here the supercurrent in the following way: i.) 
Given that
the current is fermionic and linear in the fermionic fields\footnote{This 
assumption is obviously true only if the Lagrangian is at most quadratic in the fermionic 
fields. In the case of quartic fermionic interactions such as for non-linear sigma models
the procedure proposed here has to be modified.} we make the most general ansatz 
linear in $\lambda_I$ and $\psi_i$ and their conjugates. ii.) Since the parameter $\eta_I$ of
the susy transformations (\ref{dsusy}) is a chiral symplectic Majorana-Weyl spinor the 
associated supercurrent $J^M_I$ has to be an \emph{anti}-chiral symplectic Majorana-Weyl spinor 
and thus satisfies
\begin{equation}
  \label{chJ}
    \Gamma_7\,J^M_I = - J^M_I  \  \  , \  \ J^M_I = \vep_{IJ}\, C^{-1} \bar{J}^{M\, J^{\, T}}\ .
\end{equation}
Consequently the susy transformation of fields is generated by the commutator 
\begin{equation}
  \label{dcurent}
  \delta^{\mathrm{susy}} = i\ [\,\int_{\vec{x}} \bar\eta^I J^0_I\, ,\ \ ] =  i\ [\, \bar\eta^I Q_I \, ,\ \ ]\ . 
\end{equation}
Given these preliminaries we determine the susy current and fix its ambiguities by the 
requirement that the transformation (\ref{dcurent})  generates the susy transformations 
of the fermions in (\ref{dsusy}) without surface terms, through the canonical anti-commutators 
derived from the Dirac brackets.
This requirement and Lorentz covariance gives a unique current in accordance with the transformations
of the elementary fields (\ref{dsusy}):
\begin{align}
  \label{scurrent}
  J^M_I =\,& \fr{2}{g^2}\mathrm{Tr}\,\{ \fr{1}{2}\, F_{PQ}\,\Gamma^{PQ}\Gamma^M\lambda_I 
               - i\, \vec{\aux{D}}\cdot\vec{\tau}_I{}^J\Gamma^M\lambda_J\,\} \nonumber\\[3pt]
  &\ \ +{\textstyle{\sqrt{2}}}\ \Gamma^N\Gamma^M (\, C^{-1}\bar{\psi}_i^T D_N S_{iI} - 
   \vep_{IJ} D_N \bar{S}^J_i\psi_i\,) \ . 
\end{align}
It is easy to check that this current is conserved on-shell. 
\\

\noindent
{\bf Supersymmetry Algebra.} The variation of the supercurrent under the 
susy transformations (\ref{dsusy}) gives the local susy algebra. Using the fermionic 
e.o.m.\ and the on-shell condition (\ref{d}) for the auxiliary triplet the resulting
transformation of the current (\ref{scurrent}) can be written in the form
\begin{align}
  \label{dJ}
    \delta\, J^M_I = -2\, [\,\delta_I{}^J\, (\, T^{MN} 
             - \fr{1}{4}\,\vep^{MNPQRS} \ca{Z}_{PQRS}\, )\,\Gamma_N
             +\ca{Z}_I{}^J{}_{PQ}\,\Gamma^{MPQ}\, ] \, \eta_J \ ,
\end{align}
where we did not write a fermionic topological term.\footnote{\label{ft}This term is of the form 
$\partial_P Q^{[PM]}_I$ and is thus conserved off-shell and the spatially integrated zero component gives 
a surface term in any dimension. Generally one does not expect surface contributions from fermions.
See however \cite{Mayrhofer:2007ms} for a counter example. For completeness, 
$Q^{[PM]}_I= [-\fr{4}{g^2}\mathrm{Tr}\,
               \{\bar\lambda ^J\Gamma^{[P}\eta^{M]N} \lambda_I\}\Gamma_N                
    +\fr{1}{2}\, \delta^J{}_I (\bar\psi_i\eta^{R[M}\Gamma^{P]ST}\psi_i)\Gamma_{RST}] \eta_J$. See the 
appendix for useful relations.} 
The first contribution in (\ref{dJ}) is the on-shell gravitational stress tensor
\begin{align}
  \label{TMN}
  T^{MN} =&\  \fr{2}{g^2}\mathrm{Tr}\,\{F^{MP}F^N{}_P\}\, + \, 2\, D^{(M} \bar S^K_i D^{N)} S_{iK}
   \nonumber\\ 
         & +\,\fr{1}{g^2}\mathrm{Tr}\,\{\bar\lambda^K\Gamma^{(M}D^{N)}\lambda_K\} 
          + \fr{1}{2}\,\bar\psi_i\,\Gamma^{(M}\overset{\leftrightarrow}{D}{}^{N)}\psi_i 
         + \eta^{MN}\La_{6D}\ ,
\end{align}
where the $\lambda$-term is hermitian due to the symplectic Majorana-Weyl condition and the 
fermionic part in $\La_{6D}$ vanishes on-shell.

The local topological charges, or charge densities, are given by
\begin{align}
  \label{zz}
  \ca{Z}_{PQRS} =&\  \fr{1}{g^2} \mathrm{Tr}\,\{F_{[PQ}F_{RS]}\} + \fr{1}{6}\,\partial_{[P}\,
      \big{(}\,\fr{1}{g^2} \mathrm{Tr}\,\{ \bar\lambda^K\Gamma_{QRS]}\lambda_K\} 
       - \bar\psi_i\Gamma_{QRS]}\psi_i\,\big{)}\ ,  
    \nonumber \\[8pt]
   \ca{Z}_I{}^J{}_{PQ} =&\  \vec{\tau}_I{}^J\cdot [\, \fr{i}{2}\,\vec{\xi}\ \mathrm{Tr}\,F_{PQ} + 
\partial_{[P}(\bar S_i\,\vec{\tau}\, D_{Q]} S_i)\, ]
\end{align}
The first charge  corresponds to the standard central charge of the $\sN=2$ susy algebra in four dimensions.
The fermionic terms are of the form discussed in footnote \ref{ft} and they are not 
relevant for the following discussion, see however related comments in the next section. 
In four dimensions the four-form corresponds to a point like object (by duality) just like the usual 
(un-confined) monopole
in the Coulomb phase. The second charge, however, is a two-form and thus corresponds to an extended object also 
in four dimensions.
Depending on the location in six-dimensional space these charges may describe two-branes (domain walls) or 
one-branes (strings) in four dimensions. Clearly the presence of such charges breaks four-dimensional 
Lorentz symmetry and they are therefore usually not considered in the general structure of $\sN=2$ susy-algebra. 
Equivalently one can show that these charges cannot be carried by finite energy states 
in $\mathbb{R}^{3+1}$. Putting
the system in a compact spatial volume circumvents this problem.
For the moment we continue considering the local susy algebra and
by dimensional
reduction we will see the explicit realization of the above statements.\\

\noindent
{\bf Dimensional Reduction.} Inserting the dimensional reduction rules\footnote{\label{ftac}
Note that the super 
current and thus the supercharge are \emph{anti}-chiral symplectic MW-spinors (\ref{chJ}). 
Accordingly, the explicit chiral representation of the four-dimensional spinor is of the form 
$J_I= \begin{bmatrix} i\,j_{\alpha I} \\ \vep_{IJ}\,\bar{j}^{\dot \alpha\, J}\end{bmatrix}$.} 
(\ref{6d4d}) - (\ref{A56}) into the 
zero component of (\ref{dJ}), one obtains according to equation (\ref{dcurent}) the four-dimensional local susy 
algebra in the following form:
\begin{align}
  \label{4dsusy}
  \{ Q_{\alpha I}, j^0_{\, \beta J} \} =&\  2\, \sqrt{2} \, [\ \vep_{IJ}\, \vep_{\alpha\beta}\, \ca{Z} -
          (\vec{\tau}\vep)_{IJ}\cdot \vec{\ca{T}}^{\mathrm{w}}_k\, (\sigma^k\vep)_{\alpha\beta}\ ] \ , 
      \nonumber\\[5pt]
  \{\bar Q_{\dot\beta}{}^J\, , j^0_{\, \alpha I} \} =&\ 2\, [\, \delta_I{}^J \,
        {\mathscr P}_{\mu}\, \sigma^\mu_{\alpha\dot\beta}
         + \vec{\tau}_I{}^J\cdot \vec{\ca{T}}^{\mathrm{v}}_k \, \sigma^k_{\alpha\dot\beta}\, ]\ ,
\end{align}
where the index $k=1,2,3$ runs only over the spatial components, a convention we will in general use for 
small Latin indices, and $\sigma^k$ are thus the Pauli matrices. 
Note that the matrices $(\vec{\tau}\vep)_{IJ}$ and equivalently $(\sigma^k\vep)_{\alpha\beta}$ are symmetric.

The objects on the r.h.s.\ are the following: $\mathscr{P}_\mu$ is the four-dimensional 
momentum density $T^0{}_\mu$  and the charge density $\ca{Z}$ is the usual central charge
contribution,\footnote{Chromo-electric and magnetic fields are defined as $E_k = F_{0\,k}$ and 
$B_k = \fr{1}{2}\, \vep_{kij}F_{ij}$.}
\begin{equation}
  \label{Z}
  \ca{Z} = \fr{2}{g^2}\,\del_k \mathrm{Tr}\,\{\phi^\dagger(B_k + i E_k)\} \, ,
\end{equation}
which is carried by point-like objects such as un-confined monopoles or dyons. 
The electric and magnetic field contributions
stem from $T^0{}_{5/6}$ and $\ca{Z}_{i\,j\,k\, 5/6}$ in the six-dimensional theory and are thus both carried
by pointlike objects in three spatial dimensions. The actual electric and magnetic charge of such objects 
is proportional to the imaginary and real part of $\ca{Z}$, respectively. 

The less standard terms in (\ref{4dsusy}) are the $SU(2)_R$ vector charge densities
\begin{align}
  \label{TT}
  \vec{\ca{T}}^{\mathrm{v}}_k =&\   \vec{\xi}\ \, \mathrm{Tr}\, B_k 
        - i\, \vep_{kij}\, \partial_{i} (\bar S\, \vec{\tau}\, D_j S) \ , \nonumber\\[5pt]
  \vec{\ca{T}}^{\mathrm{w}}_k =&\    \partial_k\,\big{[}\,  
                              \bar S_i\, \vec{\tau}\, (\phi^\dagger+ \bar{m}_i) S_i 
                              -  \vec{\xi}\ \, \mathrm{Tr}\, \phi^\dagger\,\big{]} \ .
\end{align}
The first charge density, $\vec{\ca{T}}^{\mathrm{v}}_k$, is carried by string like objects, such as
vortices or confined monopoles, and will be our main interest here. The second term in  
 $\vec{\ca{T}}^{\mathrm{v}}_k$ does not contribute for classical configurations but at the 
quantum level it may give interesting winding effects, as shown for the abelian 
vortex in \cite{Rebhan:2003bu}.
The second charge density, 
 $\vec{\ca{T}}^{\mathrm{w}}_k$, is carried by two-dimensional objects, i.e. domain walls. These charges
are not invariant under Lorentz and $SU(2)_R$ transformations.
For all three charge densities $\ca{Z}$, $\vec{\ca{T}}^{\mathrm{v}}_k$ and $\vec{\ca{T}}^{\mathrm{w}}_k$
we did not include the fermionic terms for reasons given above.

\subsection{$\frac{1}{4}$ BPS Multiplets}    

We discuss now the multiplet structure for states which carry the charges derived above.
Our focus is on confined monopoles and we therefore set the domain wall charge 
 $\vec{\ca{T}}^{\mathrm{w}}_k$ equal to zero in the following. To discuss the multiplet structure
we assume temporarily that the system is in a partially compactified volume so that the 
integrated charge $Z=\int_{vol}\ca{Z}$ and $\vec{T}^{\mathrm{v}}_k  = \int_{vol}\vec{\ca{T}}^{\mathrm{v}}_k$
exist; this particularly concerns the vortex charge. These $\frac{1}{4}$ BPS multiplets 
exist only due to the presence of the tensorial central charges and are different in nature
from  $\frac{1}{4}$ BPS multiplets in $\sN = 4$ SYM. 
\\

\noindent
In the following we assume that the flux associated with 
the vortices that confine the monopoles is
oriented in a single direction. Intersecting vortices are at most $\frac{1}{8}$ BPS states, 
see \cite{Eto:2005sw} for numerous aspects at the level of BPS equations. 
By rotational symmetry of (\ref{4dsusy})
the vortices can be oriented in the spatial $3$-direction, i.e. 
$\vec{T}^{\mathrm{v}}_k= (0,0,\vec{T}^{\mathrm{v}})$, which is of course the natural choice
given the conventions for the Pauli matrices. We look at representations for states 
whose rest-frames coincide with the frame of reference defined by the vortices. Thus we set
$P^\mu = \int_{vol}\mathscr{P}^{\mu}=(M,0,0,0)$. For such representations the (integrated) 
algebra (\ref{4dsusy}) can be written as
\begin{align}
  \label{bps1}
   \{ Q_{\alpha I}, Q_{\, \beta J} \} =&\  2\, \sqrt{2} \,  \vep_{IJ}\, \vep_{\alpha\beta}\, Z , 
      \nonumber\\[5pt]
  \{ Q_{\, \alpha I}\, ,\bar Q_{\dot\beta}{}^J\} =&\ 2\, [\, \delta_I{}^J \,
       M \delta_{\alpha\dot\beta}
         + \vec{\tau}_I{}^J\cdot \vec{T}^{\mathrm{v}} \, \sigma^3_{\alpha\dot\beta}\, ]\ .
\end{align}
By an $SU(2)_R$ transformation we can diagonalize the matrix 
$\vec{\tau}\cdot\vec{T}^{\mathrm{v}}$ to the form\footnote{\label{ftsu}Writing 
$\vec{\tau}\cdot\vec{T}^{\mathrm{v}}$ as 
$\begin{pmatrix} z & \bar w\\ w & -z\end{pmatrix}$ 
the transformation is given by 
$U= \frac{1}{\sqrt{2r(r-z)}}\begin{pmatrix}  w & r-z \\ -(r-z)& \bar w\end{pmatrix}$
with $r=\sqrt{z^2+|w|^2}$. This is of course none other than (one) of the $SU(2)$ 
matrices associated with the rotation which aligns the vector $\vec{T}^{\mathrm{v}}$
in the positive $3$-axis. } 
$|\vec{T}|\,\tau^3$,
with $|\vec{T}|$ being the euclidean norm of the real $SU(2)_R$ vector $\vec{T}^{\mathrm{v}}$. Introducing 
then the operators
\begin{equation}
  \label{bps2}
  \left. \begin{matrix} A_\alpha \\ B_\alpha \end{matrix}\ \right\} \ :=\  
   \frac{1}{2\sqrt{Z}}\  \big{(}\, \sqrt{\bar Z}\, Q_{\alpha\,1}\,
            \mp\, \sqrt{Z}\,  \vep_{\alpha\dot\gamma}\, \bar{Q}_{\dot\gamma}{}^2\,\big{)}, 
\end{equation}
the only non-vanishing anti-commutators are
\begin{align}
  \label{bps3}
  \{\, A_{1/2}, A^\dagger_{1/2}\} = M\pm |\vec{T}|-\sqrt{2}\,|Z|\  \ , \ \ 
  \{\, B_{1/2}, B^\dagger_{1/2}\} = M\pm |\vec{T}|+\sqrt{2}\,|Z| \ ,
\end{align}
where the upper/lower sign corresponds to the index $1$ and $2$, respectively. 
The left hand sides in (\ref{bps3}) are positive semi-definite and thus the oscillator
$A_2$ defines the BPS bound for the mass $M$. If we now choose a BPS saturated 
representation, i.e.
\begin{equation}
  \label{bps4}
  M=|\vec{T}|+\sqrt{2}\,|Z|\ ,
\end{equation}
the $A_2$ oscillator is represented trivially and one is left with a three-dimensional 
fermionic oscillator algebra. The multiplet thus has multiplicity $4+4=8$, which is a ``semi-short''
multiplet compared to a long massive multiplet with multiplicity $16$ and a short $\fr{1}{2}$
BPS multiplet with multiplicity $4$.

The BPS-spinor which generates the susy transformations that leave the states of this 
multiplet invariant is obtained from the condition
\begin{align}
  \label{bps5}
  \delta^{susy}=\ i\,\bar \eta^I\,Q_I =\  (\,\vep_{IJ}\, \bar{\epsilon}^{\,\D\alpha\,I}\bar Q_{\D\alpha}{}^{J}
                         + \vep^{IJ}\,\epsilon^{\,\alpha}{}_I Q_{\alpha\,J} \,) \ 
     \overset{!}{\sim}\  \epsilon\, A_2 + \bar\epsilon\, A_2^\dagger \, ,
\end{align}
where the first equality is the six-dimensional transformation according to (\ref{dcurent}), which is 
dimensionally reduced according to (\ref{4dchiral}) and footnote \ref{ftac}. 
This gives a condition on the transformation parameters
$\epsilon_{\alpha\,I}$ which is satisfied by the $\frac{1}{4}$ BPS spinors
\begin{equation}
  \label{bps6}
    [\,\epsilon_{\alpha\, I}\,] = \begin{bmatrix} \, 0 \  & e^{-i \vth}\,\epsilon \\ 
       \bar \epsilon & 0\,\end{bmatrix}\ \ \ \Rightarrow \ \ \  
       \bar\epsilon^{\,\D\alpha\, I}  = \, e^{i \vth}\,\vep^{IJ}\, \epsilon_{\alpha\,J}
  \, ,
\end{equation}
where we have introduced the central charge angle $\vth$ via $Z=|Z|\, e^{i\,\vth}$. The second relation,
which is implied by the first one but not the other way around, is the sole condition that one would
obtain for $\fr{1}{2}$ BPS states, i.e. when $|\vec{T}|=0$. The solution (\ref{bps6}) for the BPS spinor
has been obtained for the $SU(2)_R$ rotated case when $\vec{T}^{\mathrm{v}}$ points in the 
positive $3$-direction. The generic case is easily obtained by the inverse of the transformation given 
in footnote \ref{ftsu}. The reason for determining the BPS spinor (\ref{bps6}) is to derive
BPS equations and fermionic zero-modes. Instead of transforming the BPS spinor
we will rotate these equations back to the generic case.
\\

\subsection{$SU(2)_R$ Covariant $\frac{1}{4}$ BPS Equations.}

With the BPS spinor (\ref{bps6}) at hand one can derive the equations that a classical background has to 
satisfy for being invariant under the associated supersymmetry. For this the susy transformations of the 
fermions have to vanish. Dimensional reduction of the transformations for the fermions in (\ref{dsusy})
gives (no summation over repeated flavor indices)\footnote{Here another departure 
from Wess and Bagger conventions: $\sigma^{\mu\nu} := 
\sigma^{[\mu}\bar\sigma^{\nu]} = \fr{1}{2}(\sigma^{\mu}\bar\sigma^{\nu}-\sigma^{\nu}\bar\sigma^{\mu})$.}
\begin{align}
  \label{bpse1}
  \delta\lambda_{\alpha I}&= \big{[\,}\delta_I{}^J (\,\fr{1}{2}\sigma^{\mu\nu}{}_\alpha{}^\beta F_{\mu\nu} 
    +i\, [\,\phi,\phi^\dagger\,]\,\delta_\alpha{}^\beta\,) 
    +i \,\vec{\tau}_{I}{}^{J}\cdot\vec{\aux{D}}\, \delta_\alpha{}^\beta\,\big{]} \epsilon_{\,\beta J} 
     -i\,\sqrt{2}\,\vep_{IJ}\, \slashed{D}_{\alpha\D\beta}\, \phi\, \bar\epsilon^{\,\D\beta J}, 
     \nonumber\\[3pt]
   \delta \chi_{\alpha i}&=\,i\, \sqrt{2}\, \big{[}\, 
      \slashed{D}_{\alpha\D\beta}\, S_{i I}\, \bar\epsilon^{\,\D\beta I}
    -\sqrt{2}\, (\phi^\dagger + \bar m_i) S_{i I}\,\vep^{IJ}\epsilon_{\,\alpha J} \, \big{]},
     \nonumber\\[3pt]
   \delta\bar{\tilde{\chi}}_i{}^{\D\alpha} &=\, -\sqrt{2}\, \big{[}\, 
      \bar{\slashed{D}}{}^{\D\alpha\beta} S_{i I}\,\vep^{IJ}\,\epsilon_{\,\beta J} 
      +\sqrt{2}\,(\phi + m_i) S_{i I}\, \bar\epsilon^{\,\D\alpha I} \, \big{]}\ ,
\end{align}
where the auxiliary triplet $\vec{\aux {D}}$ is understood to be on-shell now, (\ref{d}).
The invariance of the gaugino, i.e. $\delta\lambda_{\alpha I}\overset{!}{=}0$, gives
with the  $\frac{1}{4}$ BPS spinor (\ref{bps6}) the equations
\begin{align}
  \label{bpse2}
  &\sqrt{2}\,D_0 \phi - [\,\phi,\phi^\dagger]=0\, , \nonumber\\
  &E_{k}- i\, (B_{k} - \sqrt{2}\ e^{i\,\vth}\, D_k \phi +\delta_{k 3}\, \aux{D}^3)=0\, , \nonumber\\
  &\aux{D}^1 = \aux{D}^2 =0 \, .
\end{align}
For static configurations and in temporal gauge ($A_0\equiv 0 \equiv \del_0$) the chromo-electric field
$E_k$ is identically zero and the first equation implies $[\,\phi,\phi^\dagger] = 0$. To this end one sets 
$\phi^\dagger = \phi$. Hermiticity of the second equation then implies that $ e^{i\,\vth} =\pm 1$,
hence the central charge $Z$ is real, see (\ref{Z}), and therefore purely magnetic. The equations (\ref{bpse2})
thus reduce to\footnote{\label{ftelec}There is of course a different possibility where the central charge $Z$ 
has magnetic and electric contributions. Replacing the hermitian $\phi$ of (\ref{bpse3}) by 
$e^{- i\,\vth} \ \phi$ still satisfies the first equation of (\ref{bpse2}) but the hermiticity of the 
second equation does not put any restrictions on the central charge angle $\vth$. Note that the 
electric/magnetic fields are particular combinations of the non-abelian chromo-electric/magnetic 
and adjoint scalar fields (\ref{Z}). 
Therefore the electric charge can be non-zero even though the chromo-electric field $E_k$ vanishes. The angle 
$\vth$ is thus a modulus of the classical adjoint fields, which upon quantization gives 
the electric charge for the usual, un-confined, dyon states.}
\begin{equation}
  \label{bpse3}
  B_{k} \mp \sqrt{2}\, D_k \phi +\delta_{k 3}\, \aux{D}^3 = 0 \ \ ,\ \ \aux{D}^1 = \aux{D}^2 =0 \ .
\end{equation}

\noindent
{\underline{$SU(2)_R$ Rotations:}} Before analyzing the vanishing condition for the quark transformations 
in (\ref{bpse1}) we discuss the
$SU(2)_R$ rotation to generic FI-parameters $\vec{\xi}$. The $\frac{1}{4}$ BPS spinor 
was derived for the rotated situation where $\vec{T}^{\mathrm{v}}$ is oriented in the positive $3$-direction,
see the comments below (\ref{bps1}). First we note that the second term in $\vec{T}^{\mathrm{v}}$ 
(\ref{TT}) does not contribute at the classical level, see below, so that
\begin{equation}
  \label{Tcl}
  \vec{T}^{\mathrm{v}} _{\mathrm{cl}}=   \vec{\xi}\,\int_{vol} \mathrm{Tr}\, B_{3}  
    = :\pm \, 2\pi|k^{\mathrm{vor}}| L\,  \vec{\xi}\ , 
\end{equation}
where $k^{\mathrm{vor}}$ is the winding number of the configuration and the upper/lower sign correspond to
positive/negative $k^{\mathrm{vor}}$ and $L$ is the size of the compactified dimension in the direction of the vortex. 
The positive constant $2\pi|k^{\mathrm{vor}}| L$  will play no role in
the discussion here, only the sign will enter the equations.  The 
rotation associated with the transformation given in footnote \ref{ftsu} aligns $\pm\, \vec{\xi}$ 
in the positive $3$-direction, i.e. $R_\xi\cdot \vec{\xi} = \pm\,|\vec{\xi}\,|\,(0,0,1)$.
Thus rotating the auxiliary triplet $\vec{\aux{D}}$ in (\ref{bpse2}) back to the (original) 
generic form gives
\begin{equation}
  \label{rotD}
  \vec{\aux{D}}_{\mathrm{org}} =
  R_\xi^{-1} \left(\begin{smallmatrix}0\\0\\1\end{smallmatrix}\right) \aux{D}^3
   = \pm \, \fr{\vec{\xi}}{|\vec{\xi}\,|}\, \aux{D}^3\ .
\end{equation}
Inserting $\aux{D}^3$ from the equations (\ref{bpse2}) gives the first set of the $\frac{1}{4}$ 
BPS equations for generic FI-parameters (omitting the indication ``$\mathrm{org}$''):
\begin{align}
  \label{bpsgen}
  &\sqrt{2}\,D_0 \phi - [\,\phi,\phi^\dagger]=0\, , \nonumber\\
  & E_{\alpha} - i\, (B_{\alpha} - \sqrt{2}\ e^{i\,\vth}\, D_{\alpha} \phi)=0\, ,\nonumber\\
  &\vec{\xi}\, \big{[}\, 
   E_{3}- i\, (B_{3} - \sqrt{2}\ e^{i\,\vth}\, D_3 \phi)\big{]} 
   \mp i\, |\vec{\xi}\,|\, \vec{\aux{D}}=0\, ,
\end{align}
where we have introduced indices $\alpha=1,2$ for the directions perpendicular to the 
confined monopole-vortex. In the case of static fields and using temporal gauge
one has, in analogy to before, $ e^{i\,\vth} =\pm 1$ and the  
equations reduce to 
\begin{align}
  \label{genstat}
  \vec{\xi}\,  
   (B_{3} \mp \sqrt{2}\, D_3 \phi)
   \pm \, |\vec{\xi}\,|\, \vec{\aux{D}}=0\ \ \ \ ,\ \  \ \ B_{\alpha} \mp \sqrt{2}\, D_{\alpha} \phi=0 \ ,
\end{align}
where the sign of the $\phi$ terms is uncorrelated with the sign of the  $\vec{\aux{D}}$ term. 
As shown above they
determine the signs of the charges $Z$ and $\vec{T}^{\mathrm{v}}$. In the second case it is
actually the sign of the winding number $k^{\mathrm{vor}}$
which determines if  $\vec{T}^{\mathrm{v}}$  is parallel/anti-parallel to 
$\vec{\xi}$. In both cases the upper/lower sign corresponds to 
positive/negative charge and winding number. 

The invariance condition for the quarks in (\ref{bpse1}) gives the
$\frac{1}{4}$ BPS equations for the matter multiplet. We again start first in the rotated situation 
with the  $\frac{1}{4}$ BPS spinor (\ref{bps6}). But first we analyze the $\vec{\aux{D}}$ term 
equations (\ref{bpse3}) in the rotated system. As mentioned 
$R_\xi\cdot \vec{\xi} = \pm\,|\vec{\xi}\,|\,(0,0,1)$ so that the $\vec{\aux{D}}$ 
term (\ref{d}) is rotated to
\begin{equation}
  \label{drot}
  \vec{\aux{D}} = R_\xi \cdot \vec{\aux{D}}_{\mathrm{org}} = 
    \frac{g^2}{2}\, (\vec{\tau}_I{}^J S_{i\,J}\otimes \bar S^I_i -  R_\xi \cdot\vec{\xi}\ \unit )\, ,
\end{equation}
with $S_{iI}$ being the fields transformed with $U$ given in footnote \ref{ftsu}. Therefore the 
equations $\aux{D}^{1}=0=\aux{D}^2$ of (\ref{bpse3}) and the asymptotic condition on $\aux{D}^3$  
give with (\ref{sur1})
\begin{equation}
  \label{d=0}
  q_i\otimes \tilde q_i = 0  \ \ , \ \ 
    q_i\otimes \bar q_i - \bar{\tilde q}_i\otimes \tilde q_i \mp |\vec{\xi}\, |\ \ \rightarrow \ \ 0\ , 
\end{equation}
where the second relation is the condition for finite energy (density). The exact spatial direction 
of the asymptotic is not important here, what counts is that one direction exists where this 
condition has to be met.  The equations (\ref{d=0}) are exactly of the form of the vacuum conditions
(\ref{vac}) similar to the special case i.) discussed below (\ref{vac3}). Therefore the classical squark fields
have to satisfy
\begin{align}
  \label{d=02}
  k^{\mathrm{vor}} > 0 :& \ \ \tilde{q}_i = 0 = S_{i\,2}\nonumber \\ 
  k^{\mathrm{vor}} < 0:&  \ \ q_i = 0 = S_{i\,1}
\end{align}
Since the solutions one is looking for should be topologically stable the squark fields without 
vev, and thus without any asymptotic winding, will be trivial for such solutions. Therefore  
$S_{iI}=0$ for $i>N$ in the considered vacua, see the discussion following (\ref{vac}). 

It is now convenient to describe both situations with a single set of fields $\Sigma_i$
which for the upper/lower sign (i.e. $k^{\mathrm{vor}}\gtrless 0$) is identified 
with $\Sigma_i=S_{i\,1}$ and $\Sigma_i=S_{i\,2}$.
The invariance conditions $\delta\chi_{\alpha i}=0=\delta\bar{\tilde\chi}_{i}{}^{\D\alpha}$
gives then with the  $\frac{1}{4}$ BPS spinor (\ref{bps6}),
\begin{align}
  \label{bpsS}
  &(D_1\pm i D_2)\,\Sigma_i = 0\, , \nonumber\\
  &(D_0 \pm D_3)\,\Sigma_i  -\sqrt{2}\,  e^{i\,\vth} \, (\phi+m_i)\,\Sigma_i = 0 \, , \nonumber\\
  &(D_0 \mp D_3)\, \Sigma_i + \sqrt{2} \,  e^{- i\,\vth} \, (\phi^\dagger + \bar m_i)\,\Sigma_i = 0 \, . 
\end{align}

For static configurations one finds in temporal gauge the condition 
$Im\,[\, e^{i\,\vth} \, (\phi+m_i)\,] \Sigma_i=0$, where $Im$ is the difference of 
a matrix and its hermitian conjugate. The adjoint sector already gave the condition 
$\phi^\dagger=\phi$ and $e^{i\,\vth}=\pm 1$ for static solutions. Consequently the masses
$m_i$ have to be \emph{real}\footnote{Above we mentioned the possibility to also have an electric
contribution by setting $\phi\  \rightarrow \ e^{-i\,\vth} \phi$, see footnote \ref{ftelec}. 
Including the squark equations one finds the condition 
 $m_i = e^{-i\,\vth} m_i^{\mathrm{real}}$. Put differently, for static solutions to carry 
electric charge the masses $m_i$ must have a common phase and this phase determines the electric/magnetic
contribution to $Z$.} (at least for $i\leq N$), 
a choice that we will adopt for the rest of the
article. In the static case (\ref{bpsS}) reduce to
\begin{equation}
  \label{Sstat}
  (\, D_1 \pm i\, D_2\,)\Sigma_i = 0 \ \ , \ \ \mp D_{3}\Sigma_i \pm \sqrt{2}\, (\phi+m_i)\Sigma_i = 0 \ ,
\end{equation}
where the signs of the covariant derivatives are given by the sign of the winding number $k^{\mathrm{vor}}$ and the 
other signs by the sign of the charge $Z$.

Rotating the fields back to a generic situation is now a simple matter by inserting (\ref{Tcl}) into
$U^\dagger$ as given in footnote \ref{ftsu}. Here we want to give just two special cases, 
the $D$- and $F$-vortices, 
which have been considered in the literature:
\begin{equation}
  \label{tongShif}
  \vec{\xi}_{\mathrm{org}}= \left(\begin{smallmatrix} 0\\0\\ v^2 \end{smallmatrix}\right)
    \ \ \Rightarrow \ \ S^{\mathrm{org}}_{i I} \sim \begin{bmatrix} \Sigma_i \\ 0\end{bmatrix} \ \ , \ \
    \vec{\xi}_{\mathrm{org}}= \left(\begin{smallmatrix} 0\\v^2\\ 0 \end{smallmatrix}\right)
    \ \ \Rightarrow \ \ S^{\mathrm{org}}_{i I} \sim \begin{bmatrix} 1 \\ i\end{bmatrix}  \Sigma_i  \ ,
\end{equation}
and thus in the first case one has $\tilde q_i =0$ and in the second case $q_i = \bar{\tilde q}_i$,
independent of the sign of the winding number $k^{\mathrm{vor}}$. 
\\

\noindent
It is now a simple matter to obtain the fermionic zero modes associated with the susy breaking 
from the transformations 
(\ref{bpse1}) by choosing non-vanishing components for the zero-entries of the BPS spinor
(\ref{bps6}).
\\

\noindent
{\bf Bogomolnyi Trick.} 
Having discussed in considerable detail how different choices for the FI-parameters and the 
signs of the central charges are related we now make some specific choices 
for the rest of this article. Henceforth we chose $\vec{\xi}=(0,0,v^2)$, the first case in 
(\ref{tongShif}), and we choose the classical background to have charges and winding numbers
$Z>0$ and $k^{\mathrm{vor}} > 0$. The signs for the BPS equations are selected accordingly. Furthermore 
we assume $\phi_{\,\mathrm{bkg}}$ to be hermitian and the masses $m_i$ to be real so that 
we have static solutions. For later convenience we will use calligraphic letters for
some of the classical backgrounds to distinguish them from the quantum fields, so that
the classical background solution is denoted by
\begin{align}
  \label{bkg}
   &A_k^{\mathrm{bkg}} = \ca{A}_k\ \ ,\ \ 
    D_k^{\mathrm{bkg}} = \ca{D}_k\ \ , \ \ \phi_{\,\mathrm{bkg}}=\vphi = \vphi^\dagger \, , \nonumber\\
    &S_{i1}^{\mathrm{bkg}} = \Sigma_{i} \ \ ,  \ \ S_{i2}^{\mathrm{bkg}} =0 \ \ , \ \ 
    \aux{D}^3_{\mathrm{bkg}}= \mathsf{D} \, , 
\end{align}
where for the nontrivial scalar we used the notation introduced in (\ref{bpsS}), and for
the only non-vanishing component of the classical on-shell auxiliary field (\ref{d}) 
we just omit the index. It will not appear in any other context from here on. 
 
  With this particular choice of  $\vec{\xi}$ the first description of confined monopoles in the 
given context and the associated novel BPS equations were given in \cite{Tong:2003pz}, derived using the 
Bogomolnyi trick.  For a generic static configuration of the form (\ref{bkg}) the
classical energy density in temporal gauge 
$\ca{A}_0=0$ can be written as 
\begin{align}
  \label{hstat}
  \mathcal{H^{\mathrm{stat}}} & = -\La_{4D} \nonumber\\
       &=\fr{2}{g^2}\, \mathrm{Tr}\{\fr{1}{2}\,\ca{B}_k^2 + (\ca{D}_k\,\vphi)^2 
       + \fr{1}{2}\, \aux{D}^2\} + |\ca{D}_k\Sigma_i|^2  + 
        2\, \bar\Sigma_i(\vphi+m_i)^2\Sigma_i \nonumber \\[4pt]
     & = \fr{2}{g^2}\, \mathrm{Tr}\{\fr{1}{2}\, (\ca{B}_\alpha -\sqrt{2}\, \ca{D}_\alpha \vphi)^2
            +\fr{1}2\, (\ca{B}_3 -\sqrt{2}\, \ca{D}_3 \vphi +\aux{D})^2\} \nonumber\\
        &\hspace{5mm} + |\ca{D}_z\Sigma_i|^2  +|\ca{D}_3\Sigma_i - \sqrt{2}\, (\vphi + m_i)\Sigma_i |^2 
           +\sqrt{2}\, \ca{Z} + \ca{T}^{\mathrm{v}}_3{}^3 + \sqrt{2}\, \ca{T}^{\mathrm{w}}_3{}^3 \, ,
\end{align}
where $\ca{D}_z:=\ca{D}_1+i\ca{D}_2$ and  the last three terms are the non-vanishing 
components of the charge densities (\ref{Z}), (\ref{TT})
for the given class of backgrounds (\ref{bkg}),
\begin{align}
  \label{zbkg}
  \ca{Z} &= \fr{2}{g^2}\, \del_k \mathrm{Tr} \{\vphi\,\ca{B}_k\}\ \ ,\ \ 
   \ca{T}^{\mathrm{v}}_3{}^3 = v^2\, \mathrm{Tr}\,\ca{B}_3 
    - 2\,i\, \del_{[1}(\,\bar\Sigma_i\,\ca{D}_{2]}\Sigma_i\,)\, ,
   \nonumber\\
   \ca{T}^{\mathrm{w}}_3{}^3 &= \del_{3}\,[\bar\Sigma_i(\vphi+m_i)\Sigma_i -\,v^2\,\mathrm{Tr}\,\vphi\,]\ . 
\end{align}

The squares in (\ref{hstat}) are the static $\frac{1}{4}$ BPS equations (\ref{genstat}), (\ref{Sstat}) with the 
choices for the signs and $\vec{\xi}$ described above. If they are satisfied the energy 
saturates the Bogomolnyi bound (\ref{bps4}), now including also the domain wall tension, which
we assume to be zero. 

To close this section, and as a reference point for the following, we summarize the 
static $\frac{1}{4}$ BPS equations for the above choices in a convenient way. For the hermitian
adjoint classical scalar $\vphi$ and real masses $m_i$, using (\ref{A56}), (\ref{dm}), the potential 
interaction in the second term of the last line in (\ref{hstat}) can be written in terms of the covariant 
derivative $\ca{D}_6 = \del_6 - i(\ca{A}_6 + e_{6\,i})$ where $\del_6 \equiv 0$ on the classical background. 
Introducing\footnote{The given identification corresponds to complex coordinates $z=\fr{1}{2}(x^1-i\,x^2)$.} 
besides $\ca{D}_z$ also $\ca{D}_w = \ca{D}_3+i\ca{D}_6$ the static $\frac{1}{4}$ 
BPS equations considered from here on can be written as
\begin{align}
  \label{BPS}
  &\ca{B}_k -\sqrt{2}\, \ca{D}_k\vphi + \delta_{k\,3}\, \aux{D} = 0 \quad\quad
  \textrm{with:} \ \ \aux{D}=\fr{g^2}{2}\,(\, \Sigma_i\otimes\bar{\Sigma}_i -v^2\,) \nonumber\\[3pt]
 &\ca{D}_z \Sigma_i = 0 \  , \ \ \ca{D}_w \Sigma_i = 0\, .
\end{align}
These equations, first derived and discussed in \cite{Tong:2003pz}, a priori look overdetermined but as was 
noted in \cite{Isozumi:2004vg} the equations for $k=1,2$ are identical to the integrability condition of 
the last one. This novel set of equations offers a plethora of non-trivial field configurations 
carrying various charges (\ref{zbkg}) which were analyzed in detail in \cite{Sakai:2005sp}. For us the main focus 
lies with the configurations describing confined monopoles. In \cite{Shifman:2004dr} an approximate solution 
for a single
confined monopole was given for the case of gauge group $U(1)\times SU(2)$
and $\Nf=2$ (with $\vec{\xi}$ given by the second case in (\ref{tongShif})). Equations with less supersymmetry
for intersecting vortices were derived in \cite{Eto:2005sw}.

\section{Quantization}

In this section we perturbatively quantize the theory in the $\frac{1}{4}$ BPS background of confined monopoles
and study quantum corrections for the energy and topological charges. Quantization 
in solitonic backgrounds was plagued with inconsistencies for more than a decade, even in the simplest
models, see  \cite{Rebhan:2002uk} for a detailed account. The crucial point in solving these 
inconsistencies was the development
of a consistent regularization scheme. This was achieved by embedding the solitonic objects
in a higher dimensional theory \cite{Rebhan:2002uk, Rebhan:2002yw}, with the same susy content as the 
original one, and using
the flat extra-dimensions as a dimensional regulator. In the end when removing the regulator
the extra-dimension is taken to vanish. In a sense this regularization is in the same spirit as dimensional 
regularization by dimensional reduction \cite{Siegel:1979wq} with the difference that one starts in 
a higher dimensional
space than the original theory is formulated.  This method is not only consistent but also 
surprisingly elegant and simple and could be applied also to three- and four-dimensional gauge theories and 
non-linear sigma models \cite{Rebhan:2003bu,Rebhan:2004vn,Mayrhofer:2007ms}. Another convenient side 
effect of this regularization is that 
the usual and necessarily awkward treatment of zero modes via collective coordinate quantization is 
completely absent: Through the embedding in higher dimensions zero modes become massless modes
with momenta in the flat extra dimensions and are treated on the same footing as non-zero modes.

\subsection{Fluctuation Operators, Propagators}

Following the described method we start from the 
six-dimensional setting (\ref{l6d}) and  expand the Lagrangian around 
the classical background (\ref{BPS}) to second order in quantum fields,  
which defines the associated propagators or fluctuation operators. Though the static background 
lives in strictly three spatial dimensions, for the purpose of regularization the quantum (fluctuation) 
fields are taken to depend also on the  extra dimensions. 
Hence we decompose the full quantum fields as
\begin{align}
  \label{fluc}
  A_{M} &= \ca{A}_{M} + a_{M} \quad \textrm{with:} \ \ \ca{A}_M = (\,0, \ca{A}_k,0, -\sqrt{2}\, (\vphi + m_i)\,)
   \nonumber\\
  S_{iI} &= \Sigma_{i I} + s_{iI}  \quad \ \, \textrm{with:}\,  \ \ \Sigma_{iI}= (\,\Sigma_i,0\,) \,
\end{align}
where the background fields satisfy the $\frac{1}{4}$ BPS equations (\ref{BPS}) and the quantum 
fields $a_{M}$, $s_{iI}$
depend also on the extra dimensions. The same applies to the fermions in (\ref{l6d}). 

{\bf Gauge Fixing.} Another crucial step in the successful quantization of gauge theories in the presence 
of solitons is the choice of a convenient gauge. The purpose is to diagonalize that part of the 
Lagrangian which is
quadratic in the fluctuations. The proper choice to achieve this  is a suitable generalization of the 
't Hooft $R_\zeta$-gauge, which serves the same purpose in spontaneously broken gauge theories. The 
associated gauge fixing and ghost Lagrangian is conveniently obtained by the BRS 
formalism:\footnote{For the physical fields the nil-potent BRS transformations are defined as $\lambda \delta_{\mathrm{BRS}}=\delta^{\mathrm{gauge}}_{\Lambda = \lambda c}$ with the anti-commuting parameter $\lambda$ and ghost field $c$, whereas the fixed background ``functions'' are inert. In addition, $\delta_{\mathrm{BRS}} b = B$, 
$\delta_{\mathrm{BRS}} B=0$ and $\delta_{\mathrm{BRS}} c =i\, \{c,c\}$.}
\begin{align}
  \label{gf}
  \ca{L}_{\mathrm{gf+gh}} &= \delta_{\mathrm{BRS}}\Upsilon \quad \textrm{with:}\nonumber\\[3pt]
   \Upsilon &= \fr{2}{g^2}\,\mathrm{Tr}\,\{ b\big{[}
    \ca{D}_{M}(A^M-\ca{A}^M) + \fr{i\, g^2}{2} \zeta \,( S_{iI}\otimes \bar{\Sigma}^I_i - 
     \Sigma_{iI}\otimes \bar S_i^I) +\fr{\zeta}{2} B \big{]}\} \, ,
\end{align}
where $b$ is the anti-ghost and $B$ is a (Nakanishi-Lautrup) auxiliary field. The gauge fixing Lagrangian 
$\ca{L}_{\mathrm{gf+gh}}$ is thus manifestly BRS-invariant for arbitrary background functions 
$\ca{A}_M$ and $\Sigma_{iI}$. Therefore gauge invariant quantities, e.g. the (quantum) mass
of a soliton or the gauge coupling renormalization, can be computed with a convenient choice 
for the background functions, classical BPS solutions for solitons or the vacuum for renormalization 
constants, and be compared to each other. 

Choosing the background fields in (\ref{gf}) to satisfy the $\frac{1}{4}$ BPS equations (\ref{BPS})
and integrating out the auxiliary field $B$ one obtains for the gauge-fixing and ghost Lagrangian
\begin{align}
  \label{lgfgh}
  \ca{L}_{\mathrm{gf+gh}}=& -\fr{1}{g^2}\,\mathrm{Tr}\,\{\, \fr{1}{\zeta} \big{[}\,\ca{D}_{M}\,a^M 
        +\fr{i\,g^2}{2} \zeta\, (\, s_{iI}\otimes\bar\Sigma_i^{I} - \Sigma_{iI}\otimes \bar s_i^I)\big{]}^2\}
       \nonumber\\[2pt] 
       &- \fr{2}{g^2}\,\mathrm{Tr}\,\{\,b\, \ca{D}^M D_{M} c 
   - \fr{g^2}{2} \zeta\,\big{[} \, b\,( \Sigma_{iI}\otimes\bar S_i^I)\, c 
                                 -  c\, (S_{iI}\otimes\bar\Sigma_i^I\,)b\,\big{]}\}\, ,
\end{align}
where we made use of the decomposition (\ref{fluc}). In the following the choice $\zeta =1$ will lead to
particular simplifications, namely diagonalized kinetic terms. 
\\

\noindent
{\bf Fluctuation Lagrangian.} The next step is to expand the Lagrangian (\ref{l6d}) and the 
gauge fixing Lagrangian (\ref{lgfgh}), both a priori in six dimensions, to second order in the 
quantum (fluctuation) fields to obtain the fluctuation operators which define the propagators.

We start with the fermionic part of  (\ref{l6d}). The particular form of the BPS background in 
(\ref{fluc}) and the need for a  proper regularization through a flat extra dimension leads to a 
particular choice for the representation of the four-dimensional gamma matrices in (\ref{6d4d}),
which will differ from the chiral representation (\ref{4dchiral}). 

In second order of the quantum fields the fermionic part of (\ref{l6d}) for the background fields 
as given in (\ref{fluc}) reads  
\begin{equation}
  \label{lf2}
  \ca{L}_{\mathrm{ferm}}^{(2)} = -\fr{2}{g^2}\,\mathrm{Tr}\,\{\bar\lambda^2\,\Gamma^M\ca{D}_{M}\lambda_2\}
       -\bar\psi_i\, \Gamma^M\ca{D}_{M}\psi_i 
    + i\,\sqrt{2}\, (\, \bar\Sigma_i\,\bar\lambda^2\,\psi_i - \bar\psi_i\,\lambda_2\,\Sigma_i\,) , 
\end{equation}
where we have used the symplectic MW-condition (\ref{6dspin}) to express\footnote{\label{ftst}We skipped here the 
fermionic surface term $-\fr{1}{2}\del_{M}(\bar\lambda^1\Gamma^M\lambda_1)$ which has the same 
standard form as for the $\psi_i$ kinetic Lagrangian to make it hermitian.} $\lambda_1$ in terms of 
$\lambda_2$ which couples to the background scalars $\Sigma_i$. Inserting the background gauge field
(\ref{fluc}) and using the decompositions (\ref{6d4d}), (\ref{6d4dspin}) for six-dimensional spinors 
and gamma matrices the Dirac operator on $\lambda_2$ in terms of four-component spinors reads
\begin{equation}
  \label{dl}
  \Gamma^M\ca{D}_M\lambda_2 = 
    \big{[}\,(\gamma^0\del_0+\gamma_5 \del_5) + \gamma^k\ca{D}_k 
  + i\,\unit (\del_6 + i\,\sqrt{2}\, [ \vphi,\, .\,]) \, \big{]}\lambda\otimes \textstyle{\binom{0}{1}}\ ,
\end{equation}
where we omit the index for the four-component spinor $\lambda$. A similar structure 
is given for the quarks $\psi_i$.
One now has to select the regulating dimension and appropriate representation for $\gamma^\mu$:
i.) The regulating flat extra dimension should be in the same matrix block as the already present flat 
temporal direction. ii.) Both flat directions, time and regulator, should be in the opposite matrix
block to the background fields. This allows a simple diagonalization of the resulting fluctuation 
operators. From the last term in (\ref{dl}) one sees that the diagonal matrix blocks are already 
occupied by the background field $\vphi$. Therefore we keep the $x^5$ direction as regulating dimension,
i.e.\ \emph{all} fields are taken to be independent of $x^6$ in the following, and we choose the 
representation for the Dirac matrices $\gamma^\mu$ accordingly. Consequently we have, 
\begin{equation}
  \label{4dgrep} 
  \del_6 \equiv 0\ \ \  \textrm{and}\ \ \  
  \gamma^k  =  \begin{pmatrix} \sigma^k & 0 \\ 0 & - \sigma^k \end{pmatrix}\ , \ \
   \gamma^0 =  \begin{pmatrix}\, 0\, & \,i\, \\ \,i\, & \,0\, \end{pmatrix}\ , \ \
   \gamma_5 =  \begin{pmatrix} 0 & -i \\ i &\, 0 \end{pmatrix}\ , 
\end{equation}
where each entry is a two by two matrix ($\sigma^k$ the Pauli matrices) and $\gamma_5$ is obtained 
according to the definition given below (\ref{6d4d}). 

To write the quadratic fermionic Lagrangian (\ref{lf2}) in a convenient way we introduce the 
following structures: First we group together the nontrivial vector field backgrounds
\begin{equation}
  \label{A4}
   \left.
  \begin{array}{l} 
     \ca{A}_{\ww k} = (\,\ca{A}_k, -\sqrt{2}( \vphi + m_i)\,) \\ \ \del_{\ww k}=(\, \del_k,\, 0)
   \end{array}\right\} \quad
    \ca{D}_{\ww k} = \del_{\ww k} -i\, \ca{A}_{\ww k} \, 
\end{equation}
where the masses $m_i$  act only on the quarks $\psi_i$. Secondly we introduce the Euclidean quaternions
according to the block decomposition (\ref{4dgrep}),
\begin{align}
  \label{euk}
  \sigma^{\ww k} = &\ (\,\sigma^k, i\unit_2 \,)\ \ \ \ \, , \ \ 
     \slashed{\ca{D}} = \sigma^{\ww k}\, \ca{D}_{\ww k} \  \ , \ \ 
   \sigma^{\ww k \ww l} = \sigma^{[\,\ww k}\bar\sigma^{\ww l\,]}\ \ \ldots \ \ \textrm{self dual}
   \nonumber\\
  \bar\sigma^{\ww k} =& \ (\,\sigma^k, - i\unit_2 \,)\ \ , \ \ 
      \bar{\slashed{\ca{D}}} = \bar\sigma^{\ww k}\, \ca{D}_{\ww k} \ \ , \ \ 
   \bar{\sigma}^{\ww k \ww l} = \bar\sigma^{[\,\ww k}\sigma^{\ww l\,]}\ \ \ldots \ \ \textrm{anti-self dual}.
\end{align}
Decomposing also the four-component fermions $\lambda$ and $\psi_i$ according to the two by two blocks
in (\ref{4dgrep}),
\begin{equation}
  \label{fpm}
  \lambda= \left( {\lambda_{+} \atop\lambda_{-}}\right) \quad , \quad 
   \psi_i = \left( {\psi_{i\,+} \atop \psi_{i\, -}}\right) \, ,
\end{equation}
the quadratic fermionic Lagrangian (\ref{lf2}) can be written in the convenient form
\begin{equation}
  \label{lf22}
  \ca{L}_{\mathrm{ferm}}^{(2)} = \mathrm{Tr}\,\{\, U^\dagger (\, i\, \del_{+} U + L^\dagger V\,) +  
   V^\dagger (\, i\, \del_{-} V +L\, U)\,\}\ ,  
\end{equation}
where $\del_{\pm}=\del_0\pm\del_5$ and the four-component objects $U$ and $V$ are defined as
\begin{equation}
  \label{UV}
  U = \begin{bmatrix}
      \fr{\sqrt{2}}{g}\, \lambda_{+} \\ \psi_{i\,+}
     \end{bmatrix} \quad , \quad
  V  = \begin{bmatrix}
      \fr{\sqrt{2}}{g}\, \lambda_{-} \\ \psi_{i\,-}
     \end{bmatrix} \ ,
\end{equation}
i.e.\ they are mixtures of adjoint and fundamental fields, and the trace operation
in (\ref{lf22}) is defined accordingly. The most important quantities in (\ref{lf22}) are the 
fluctuation operators 
$L$ and $L^\dagger$. They are given by
\begin{equation}
  \label{LLd}
  L= \left[
     \begin{array}{cc}
     \slashed{\ca{D}}^{\mathrm{a}}  &  - ig\, \bar{\Sigma}_i^{\,\mathrm{r}} \\[3pt]
     ig\, \Sigma_i^{\,\mathrm{r}}   &  \bar{\slashed{\ca{D}}}{}^{\mathrm{f}}
     \end{array} \right] 
      \quad \ , \quad 
  L^\dagger= \left[
     \begin{array}{cc}
     -\bar{\slashed{\ca{D}}}^{\mathrm{a}}  &  - ig\, \bar{\Sigma}_i^{\,\mathrm{r}} \\[3pt]
     ig\, \Sigma_i^{\,\mathrm{r}}   &  - \slashed{\ca{D}}{}^{\mathrm{f}}
     \end{array} \right] \, , 
\end{equation}
where the superscripts $\mathrm{a}, \mathrm{f}$  indicate the adjoint and fundamental action, whereas
the superscript $\mathrm{r}$ indicates action from the right. Explicitly, the right action on an adjoint
field $X$ is just matrix multiplication, $\Sigma_i^{\, \mathrm{r}} \cdot X := X\Sigma_i$, and on a 
fundamental field $y_i$ it is tensor multiplication as used already several times 
(see for example (\ref{aux})),
$\bar{\Sigma}_i^{\, \mathrm{r}} \cdot y_i := y_i\otimes\bar\Sigma_{i}$. These rules are somehow obvious from 
the representation in which $\Sigma_i, \bar\Sigma_i$ live. As usual, summation over repeated flavor indices 
is  implied. 

The operators act in the space of the direct sum of adjoint and fundamental fields, for example 
$L:\left[{X\atop y_i}\right] \rightarrow \left[{X'\atop y'_i}\right]$. It is with regard to the natural scalar 
product in this space that the formally adjoint operator $L^\dagger$ is the hermitian conjugate of $L$, 
and vice versa. Comparing these operators 
with the fluctuation operators for the Coulomb phase  monopole and the abelian vortex 
\cite{Rebhan:2003bu, Rebhan:2006fg} 
they, naturally, resemble a combination of both structures.

Before analyzing the quadratic bosonic and gauge fixing Lagrangian we give the products of 
the operators, which will be a useful input for these and further considerations:
\begin{align}
  \label{LL}
   L^\dagger L=& \left[
     \begin{array}{cc}
     - \bar{\slashed{\ca{D}}}\slashed{\ca{D}}^{\mathrm{a}} +g^2 (\Sigma_k\otimes\bar\Sigma_k)^{\mathrm{r}}  &   
      - ig\,(\bar{\slashed{\ca{D}}} \bar{\Sigma}_j)^{\,\mathrm{r}} \\[3.5pt]
     ig\, (\slashed{\ca{D}}\Sigma_i)^{\,\mathrm{r}}   &  
       -\delta_{ij}\slashed{\ca{D}}\bar{\slashed{\ca{D}}}{}^{\mathrm{f}} + g^2\,\bar\Sigma_j\Sigma_i
     \end{array} \right] 
      \ , \nonumber\\[8pt]
   L L^\dagger=& \left[
     \begin{array}{cc}
     - \slashed{\ca{D}}\bar{\slashed{\ca{D}}}^{\mathrm{a}} +g^2 (\Sigma_k\otimes\bar\Sigma_k)^{\mathrm{r}}  &   
      ig\,(\slashed{\ca{D}} \bar{\Sigma}_j)^{\,\mathrm{r}} \\[3.5pt]
     -ig\, (\bar{\slashed{\ca{D}}}\Sigma_i)^{\,\mathrm{r}}   &  
       -\delta_{ij}\bar{\slashed{\ca{D}}}\slashed{\ca{D}}{}^{\mathrm{f}} + g^2\,\bar\Sigma_j\Sigma_i
     \end{array} \right]\ , 
\end{align}
which now also have a nontrivial matrix structure in flavor space and act as 
$\left[{X\atop y_j}\right] \rightarrow \left[{X'\atop y'_i}\right]$.

Expanding the bosonic part of the Lagrangian (\ref{l6d}) to second order in the quantum (fluctuation)
fields (\ref{fluc}) and also the gauge fixing Lagrangian (the first term in (\ref{lgfgh})) the mixed 
kinetic terms between the $a_M$ and $s_{iI}$ fluctuations cancel for $\zeta =1$. The total quadratic
bosonic Lagrangian then becomes
\begin{align}
  \label{lb2}
  \ca{L}_{\mathrm{bos}}^{(2)} =&\ \fr{2}{g^2}\,\mathrm{Tr}\,\{
   - \fr{1}{2} \left[ {a^0\atop a^5}\right] \big{(}\, \del_{+}\del_{-} - \ca{D}_{\ww m}^2
  + g^2\, \Sigma_i\otimes\bar\Sigma_i\,\big{)}\left[ {a_0\atop a_5}\right] \nonumber\\[4pt]
 &\hspace{1cm} -\fr{1}{2}\, a_{\ww k}\, \big{[}\, \delta_{\ww k\ww l}\, (\, \del_{+}\del_{-} -\ca{D}_{\ww m}^2 
    +  g^2\, \Sigma_i\otimes\bar\Sigma_i\,) + 4 i\, \ca{F}_{\ww k\ww l}\, \big{]}\, a_{\ww l}\, \} \nonumber\\[4pt]
  &\hspace{-1cm} - \left[ {{\bar s_i^1} \atop {\bar s_i^2}}\right]\, \big{[}\, 
      \delta_{ij}\,(\, \del_{+}\del_{-} - \ca{D}_{\ww m}^2\,) 
       + \bar\Sigma_j\Sigma_i + \aux{D}\,\tau^3 \,\big{]} \left[ {s_{j\,1}\atop s_{j\,2}}\right]
   +2i\, \big{(}\, \ca{D}_{\ww k}\bar\Sigma_ia^{\ww k} s_{i1} - \bar s_i^1 a^{\ww k}\ca{D}_{\ww k} \Sigma_i \,\big{)},
\end{align}
where we skipped a number of total derivative terms (we will comment on this below) which however 
do not enter in the 
definition of the propagators. The $a_0,a_5$ fluctuations are already diagonal, the $a_{\ww k}$ fluctuations
of the vector field in the directions occupied by the nontrivial background (\ref{BPS}) have an additional 
spin coupling to the background field. The last line gives the fluctuations for the squark scalars, the 
first term in matrix notation w.r.t.\ the $SU(2)_R$ space with $\tau^3$ being the third $SU(2)_R$  Pauli matrix.

The aim is now to write these fluctuations in terms of the operators $L, L^\dagger$ to make
the symmetry between fermions and bosons, i.e.\ the unbroken susy in the $\fr{1}{4}$ BPS background,
manifest and exploit it in computing quantum corrections. To this end we first give some 
properties of the building blocks of the operators (\ref{LLd}), (\ref{LL}) in 
$\fr{1}{4}$ BPS background (\ref{BPS}):
\begin{align}
  \label{ddbps}
  &\bar{\slashed{\ca{D}}}\slashed{\ca{D}} = \ca{D}_{\ww k}^2 - \sigma^3\,  \aux{D}  \ \ , \ \ 
    \hspace{2.5cm}
    \bar{\slashed{\ca{D}}} \Sigma_i = 
   \begin{pmatrix} \ca{D}_{\bar w} & \ca{D}_{\bar z} \\ 0 & 0 \end{pmatrix}\Sigma_i\ ,
  \nonumber\\
  & \slashed{\ca{D}}\bar{\slashed{\ca{D}}} = 
   \ca{D}_{\ww k}^2 +\sigma^k\, (\, 2\,\ca{B}_k +\delta_{k,3}\aux{D}\,) \ \ , \ \ \ \,
    \slashed{\ca{D}}\bar\Sigma_i =
   \begin{pmatrix} \ca{D}_{ w} & 0\\ \ca{D}_{ z} & 0 \end{pmatrix}\bar\Sigma_i\ ,
   \nonumber\\
  & \ca{F}_{\ww k\ww l}^{(+)} = -\fr{i}{2}\, \aux{D}\begin{bmatrix} \sigma^2 & 0\\ 0 & \sigma^2\end{bmatrix} \ ,
\end{align}
where $\ca{F}_{\ww k\ww l}^{(+)}$ is the self dual part of $\ca{F}_{\ww k\ww l}= i\, [\ca{D}_{\ww k},\ca{D}_{\ww l}]$ .
With these relations we can express the quadratic bosonic Lagrangian (\ref{lb2}) in the simple form
\begin{align}
  \label{lb22}
   \ca{L}_{\mathrm{bos}}^{(2)} =&\ \fr{2}{g^2}\,\mathrm{Tr}\,\{
   - \fr{1}{2} \left[ {a^0\atop a^5}\right] \big{(}\, \del_{+}\del_{-} - \ca{D}_{\ww m}^2
  + g^2\, \Sigma_i\otimes\bar\Sigma_i\,\big{)}\left[ {a_0\atop a_5}\right]\} \nonumber\\[4pt]
  &\ \ - \mathrm{Tr}\,\{ W^\dagger \, \big{(}\, \del_{+}\del_{-} + L L^\dagger\,\big{)}\,W\,\} ,
\end{align}
where we again omitted a total derivative term and the four-component field $W$ is now defined as
\begin{equation}
  \label{W}
  W = [\ \fr{1}{g}\, a_w\, ,\ \fr{1}{g}\, a_z\, ,\ s_{i\,1}\,,\ s_{i\,2}\ ]^T \, ,
\end{equation}
in analogy to $U, V$ (\ref{UV}). A curious fact is that to bring the bosonic quadratic Lagrangian into
this form one has to identify the $SU(2)_R$ Pauli matrix $\tau^3$ with the space-time Pauli matrix 
$\sigma^3$ as can be seen from the second last term in (\ref{lb2}) and the first relation in (\ref{ddbps}).
Though spatial and $SU(2)_R$ orientations are completely uncorrelated at the level of the classical 
equations (\ref{genstat}), it seems that at the quantum level supersymmetry
links the spatial $SU(2)$ with the $R$-symmetry $SU(2)$.

The last missing piece is the ghost Lagrangian in (\ref{lgfgh}) in second order of the quantum fields, 
which for $\zeta=1$ is simply
\begin{equation}
  \label{lg2}
  \ca{L}_{\mathrm{gh}}^{(2)} =  \fr{2}{g^2}\,\mathrm{Tr}\,\{\,
  \fr{1}{2}\, b\, \big{(}\, \del_{+}\del_{-} - \ca{D}_{\ww m}^2
  + g^2\, \Sigma_i\otimes\bar\Sigma_i\,\big{)}\, c -(\, b\leftrightarrow c\,)\} \, ,
\end{equation}
and is thus of the same form as the $a_0, a_5$ fluctuations. As already observed for the Coulomb monopole 
they form a quartet with the ghosts $b, c$ \cite{Rebhan:2006fg}.

Finally, for completeness, we give here the total derivative terms that were mentioned above. They
are
\begin{align}
  \label{totd}
  \ca{L}_{\mathrm{tot. der.}} =& 
     - \del_M \big{[}\, \fr{2}{g^2}\,\mathrm{Tr}\,\{a^M\ca{D}_N a^N + b\,\ca{D}^Mc\}
   +i\,(\, \bar\Sigma_i\,a^Ms_{i\,1} - 
       \bar s^1_i\, a^M\Sigma_i\,) \big{]} \nonumber\\[3pt]
     &\ \  -\del_M \,\mathrm{Tr}\,\{W^\dagger \ca{D}^M W\} 
     + \fr{1}{g^2} \mathrm{Tr}\,\{ \del_M\del_N\, (\,a^M a^N\,)+ \del^2(\,b\,c\,)\}\, , 
\end{align}
where we grouped the terms in a way that will become rather convenient in a moment. It will turn 
out that the first line is BRS-exact.
These total derivative terms are not taken into account for the definition of the fluctuation
operators, or equivalently propagators.
 They may however contribute in the form of composite operator
renormalization for the energy/Hamiltonian. The same applies to the fermionic surface terms which
were mentioned in the previous section.\footnote{Basically these terms are of the same origin as the fermionic 
total derivative term mentioned in footnote \ref{ftst}. They appear because the (quadratic) 
Lagrangian given here is 
not written  in a manifest hermitian form, but hermitian only up to total derivatives. 
This is of course very standard in ordinary QFT in the vacuum sector. The difference here is that in the 
soliton sector in particular cases such surface terms might cause a composite operator renormalization for 
the Hamiltonian and central charges which is related to the difference between the ordinary current 
multiplet and the improved current multiplet. In the Coulomb phase this difference affects only conformal 
theories like $\sN=4$ SYM. However, in the present case the behavior at the boundary is rather peculiar as we 
discuss below.} 
 We will not consider such issues here, see \cite{Rebhan:2006fg,Rebhan:2005yi}
for a detailed discussion for the case of the monopole in the Coulomb phase, but leave it for a future 
analysis.
\\

\noindent
{\bf Fluctuation Equations, Propagators.} To (perturbatively) quantize the theory in the 
given $\fr{1}{4}$ BPS background one has to solve the field operator equations. 
It is sufficient to study the fermionic 
system (\ref{lf22}) since it contains all the information needed for the bosonic sector, as we will see. 
The linearized field equations obtained from 
the quadratic Lagrangian (\ref{lf22}) are
\begin{align}
  \label{fmod1}
  \begin{array}{c}
  \ L\, U = -i\,\del_{-}\, V \\
  L^\dagger V = -i\, \del_{+}\, U \end{array} \quad \quad \Rightarrow \quad\quad 
  \begin{array}{c}
   L^\dagger L\ U = - \del_{+}\del_{-}\, U \\
   \ L L^\dagger  V = - \del_{+}\del_{-}\,  V \end{array} \,
\end{align}
where we just iterated the first set of equations. The next step is to decompose the quantum fields $U, V$
into eigenmodes of $L^\dagger L$ and $LL^\dagger$. With the individual modes  normalized as  
\begin{equation}
  \label{fmod2}
  U_n^{(\pm)} = \pm\sqrt{E_n+\ell}\ u_n\, e^{\pm i(E_n\,t -\ell\, x^5)}\quad , \quad 
  V_n^{(\pm)} = \sqrt{E_n-\ell}\ v_n\, e^{\pm i(E_n\,t -\ell\, x^5)} \ ,
\end{equation}
where $E_n^2 = \omega_n^2 +\ell^2$, the field equations (\ref{fmod1}) are equivalent 
to the following canonically normalized susy quantum mechanical system:
\begin{align}
  \label{fmod3}
  L^\dagger L\ u_n& = \omega_n^2\, u_n \quad \quad\quad \quad
   v_n = \frac{1}{\omega_n}\, L\ u_n \nonumber\\[2pt]
  \, LL^\dagger\, v_n  &= \omega_n^2\, v_n  \quad \quad\quad \quad
   \, u_n = \frac{1}{\omega_n}\, L^\dagger\, v_n  \, 
\end{align}
where the algebraic relation between $u_n, v_n$ holds only for $ \omega_n\neq 0$. Thus for non-zero
modes, 
$ \omega_n\neq 0$, the modes of the $U$ and $V$ field are isospectral, though in general with different 
spectral densities for the continuum modes. The quantum number $n$ is an abbreviation for
all quantum numbers given by the operators $L^\dagger L$, $LL^\dagger$. Thus in the following we 
understand $n=\{n_s\}$, where $\{n_s\}$ is a finite set of quantum numbers for the modes $u_n, v_n$. 
These quantum numbers will consist of discrete ones and, for the continuum states, also 
continuous (momentum) quantum numbers. For zero modes of the susy quantum mechanical system (\ref{fmod3})
the mode energy is $E_n=|\ell|$ and from (\ref{fmod2}) one sees that they are massless modes 
of the quantum fields $U, V$ propagating in the flat extra dimension $x^5$. The normalization 
factor in (\ref{fmod2}) seems to be problematic for the zero-modes since it vanishes for positive
or negative momentum $\ell$ for either one or the other zero modes. 
We will show that there is no way around this, on the contrary, it has an important physical effect.
Otherwise the zero modes are treated
on the same footing as non-zero modes in the full quantum fields. 

The modes (\ref{fmod3})  are ortho-normalized
as $\int d^3 x\ \mathrm{Tr}\,\{\, u_n^\dagger\, u_m\,\} = \delta(n-m)$, and similarly for the $v$-modes, where $\delta(n-m)$ is 
a product of Kronecker deltas, for the discrete quantum numbers in $n=\{n_s\}$, and Dirac deltas
for the continuous ones. The completeness relation for the modes reads
\begin{equation}
  \label{comp}
  \sum_{n^{\mathrm{u}}_0} u_{n^{\mathrm{u}}_0}(x) u^\dagger_{n^{\mathrm{u}}_0}(x') 
     +  \int_n\!\!\!\!\!\!\!{\textstyle{\sum}}\, \mu_n\, u_n(x)u_n^\dagger(x') 
    = \delta^3(x-x')\ \unit\, ,
\end{equation}
where we introduced the symbol $\int\!\!\!\!\!\!{\scriptstyle{\sum}}_n$ to indicate summation 
over discrete quantum numbers and integration with a measure factor $\mu_n$ over continuum modes.
This measure factor will be determined below. The very same relation holds for the $v$-modes.
We indicated with  ${n^{\mathrm{u}}_0}, {n^{\mathrm{v}}_0}$  the zero modes of $L$, 
i.e. $L\,u_{n^{\mathrm{u}}_0}=0$  and $L^\dagger$, i.e. $L^\dagger v_{n^{\mathrm{v}}_0}=0$ which are not 
related to each other.\footnote{A simple argument for the norm of zero modes also shows that 
$L^\dagger L \psi_0=0 \Leftrightarrow L\psi_0=0$ and $LL^\dagger \psi_0=0 \Leftrightarrow L^\dagger\psi_0=0$.}
Usually one of the two operators does not have zero modes. The 
situation here is different as we will discuss below.

The fermionic quantum field then reads
\begin{align}
  \label{fqf}
  \Psi := \left[ {U \atop V}\right] =&\int \frac{d^{\epsilon}\ell}{(2\pi)^{\epsilon/2}} 
    \sum_{n_0^{\mathrm{u}}} \frac{1}{\sqrt{2|\ell|}} \, \Big{(}\  b_{n_0\,\ell}^{\mathrm{u}}
    \left[ {U_{n^{\mathrm{u}}_0}^{(-)} \atop 0}\right] + d_{n_0\,\ell}^{\mathrm{u}\,\dagger} 
      \left[{ U_{n^{\mathrm{u}}_0}^{(+)} \atop 0}\right]\ \Big{)} +(\,U \rightarrow \,V) \nonumber\\[4pt]
 &+ \int \frac{d^{\epsilon}\ell}{(2\pi)^{\epsilon/2}} 
     \int_n\!\!\!\!\!\!\!{\textstyle{\sum}}\sqrt{\mu_n}\ \frac{1}{\sqrt{2E_n}}\, \Big{(}\ 
      b_{n\,\ell} \left[ {U_{n}^{(-)} \atop  V_{n}^{(-)}}\right] +  d_{n\,\ell}^\dagger
    \left[{ U_{n}^{(+)} \atop  V_{n}^{(+)}}\right]\  \Big{)}\, ,
\end{align}
where $(\,U \rightarrow \,V)$ indicates a copy of the first term with obvious changes. The integral 
$\int d^{\epsilon}\ell$ is an $\epsilon$-dimensional integral from dimensional regularization. 
In the end the regulator $\epsilon$ is taken to be zero as $\epsilon \rightarrow 0^+$. All oscillator 
operators satisfy the anti-commutator relations
\begin{equation}
  \label{ac}
  \{\Psi(x,t)\, , \Psi^\dagger(x',t)\} =\delta^{3+\epsilon}(x-x') \unit\quad \Rightarrow \quad 
    \{ b_A, b_B^\dagger\}=\delta_{AB} = \{d_A, d_B^\dagger\}
   \, ,
\end{equation}
with $A, B$ being abbreviations for the super/sub-scripts appearing in (\ref{fqf}). From the normalization 
in (\ref{fmod2}) one sees that $u_{n^{\mathrm{u}}_0}$ particles can be created only for momenta $\ell<0$,
whereas $v_{n^{\mathrm{v}}_0}$ particles can be created only with momentum $\ell>0$. Therefore
the zero-modes introduce a current of a certain chirality in the extra dimension. A mismatch 
in the number of $u$ and $v$-zero modes leads to a non-vanishing current in the extra-dimension
which is related to anomalies \cite{Rebhan:2002yw,Rebhan:2004vn}.\footnote{Writing the zero 
modes without the normalization factors
that make them chiral one needs a wave-function $e^{-i\ell(t-x^5)}$ to satisfy the field equations, which 
leads to the same conclusion since particle creation operators become anti-particle creation operators
for negative momentum.}

In the following it will be less cumbersome to write down the propagators directly. Before we do this 
for the fermions, for practical reasons,  we give propagators for the bosons and the ghost.
\\

\noindent
{\underline{$W$-Bosons:}} From (\ref{lb22}) one sees that the $W$-bosons satisfy the same 
field equation as the $V$-field in (\ref{fmod1}). The propagator is thus given by the 
$v$-modes,
\begin{align}
  \label{propW}
 & \Delta_{LL^\dagger}(x,x') :=   \langle\, W(x)\, W^\dagger(x')\, \rangle  \nonumber\\[5pt]
   &\hspace{1cm} 
   =  -i \int \frac{dp^{1+\epsilon}}{(2\pi)^{1+\epsilon}}\ e^{ip_\alpha(x-x')^\alpha}\ \Big{\{} 
      \sum_{n_0^{\mathrm{v}}} \frac{v_{n_0^{\mathrm{v}}}(x) v^\dagger_{n_0^{\mathrm{v}}}(x')}{p^2-i\epsilon} 
   + \int_n\!\!\!\!\!\!\!{\textstyle{\sum}}\, \sqrt{\mu_n}\ \, 
   \frac{ v_{n}(x) v^\dagger_{n}(x')}{p^2+\omega_n^2-i\epsilon}\ \Big{\}}\, ,
\end{align}
where $ \langle\, W(x)\, W^\dagger(x')\, \rangle$ implies the time-ordered product and  
we introduced the two-momentum $p^\alpha = (p^0,\ell)$. 
In the very same form one can  write down a propagator $\Delta_{L^\dagger L}(x,x')$ by replacing the 
$v$-modes by the $u$-modes. 
\\

\noindent
{\underline{Quartet:}} The quartet $(a_0,\, a_5,\, b,\, c,)$ is governed by the same fluctuation 
operator, see (\ref{lb22}) and (\ref{lg2}). In fact they are also related to the $L^\dagger L$ operator.
With the properties (\ref{ddbps}) in the BPS-background the $(1,1)$-component of the matrix $L^\dagger L$
(\ref{LL}) decouples from the rest of the operator and acts on an adjoint field $X$ as
\begin{equation}
  \label{LL11}
  (L^\dagger L)^{1\,1}\cdot X = - \ca{D}_{k}^2\, X +\fr{g^2}{2} \{\Sigma_i\otimes\bar\Sigma_i, X\}\, ,
\end{equation}
where the last term is the anti-commutator of the two matrices. Using the cyclicity of the 
trace the last term in the fluctuation operator for the quartet 
$(a_0,\, a_5,\, b,\, c,)$ can be brought into the same form and thus it satisfies the 
same equation as the decoupled $U^1$-component in (\ref{fmod1}). One thus has the propagators
\begin{equation}
  \label{propQ}
  \langle\, a_5(x)\, a_5(x')\, \rangle = 
  - \langle\, a_0(x)\, a_0(x')\, \rangle = 
  \langle\, b(x)\, c(x')\, \rangle = \Delta^{(1\,1)}_{L^\dagger L}(x,x')\ ,
\end{equation}
where $\Delta^{(1\,1)}_{L^\dagger L}(x,x')$ is the $(1,1)$-component of the propagator of the form
(\ref{propW}) with the $v$-modes replaced by the $u$-modes. The signs are determined by 
the sign in the Lagrangian of the respective terms and the statistics of the field.
\\

\noindent
{\underline{Fermions:}} Noticing that
\begin{equation}
  \label{HH}
  \begin{bmatrix} i\del_+ & L^\dagger \\ L & i\del_- \end{bmatrix}\ 
    \begin{bmatrix} -i\del_+ & L^\dagger \\ L & -i\del_- \end{bmatrix} = 
    \begin{bmatrix} \del_+\del_- + L^\dagger L &  \\  & \del_+  \del_- + LL^\dagger \end{bmatrix}\ \ ,
\end{equation}
where the first factor is the fluctuation operator for $\Psi =(U,V)$, the fermionic propagator 
is conveniently computed as
\begin{equation}
  \label{propF}
   \langle\, \Psi(x)\, \Psi^\dagger(x')\, \rangle  =  
     \begin{bmatrix} -i\del_+ & L^\dagger \\ L & -i\del_- \end{bmatrix}
      \begin{bmatrix}  - \Delta_{L^\dagger L}(x,x') & \\ & - \Delta_{L L^\dagger }(x,x') \end{bmatrix}\ .
\end{equation}

These are the building blocks for computing quantum corrections in the BPS-background, which we discuss 
in the next section.

\subsection{Energy Correction and Anomaly}

In this section we use the results derived above to compute quantum corrections
for confined monopoles. In doing this one needs some knowledge of their classical properties.
\\

\noindent
{\bf Classical Asymptotics and Energy.}\label{asymbeh} So far we did not specify the 
classical background except for the fact that it satisfies the $\frac{1}{4}$ BPS equations 
(\ref{BPS}) and thus has an axial orientation in the $x^3$-direction. Otherwise
the above considerations are completely general. The field configuration that we have in mind in the
following is that of (multiple) confined monopoles, as depicted  
in figure \ref{fig}.\footnote{Actually, the calculations in the following include automatically
also non-abelian vortices, but the most interesting results will be due to confined monopoles.}

The axial orientation of the field configurations implies that the asymptotic boundary has the form of an 
infinite cylinder, see figure \ref{fig}, and accordingly we have to specify the asymptotic behavior: 
i.) At $x^3 \rightarrow \pm\infty$ the boundary is given by the infinite discs $\mathfrak{D}_{\pm}$
at which the fields behave like (multiple) vortices, though in general different vortices at 
$\mathfrak{D}_{+}$ and $\mathfrak{D}_{-}$. ii.) For $r \rightarrow \infty$, with $r$ being the 
radial cylindrical coordinate, the boundary is given by the cylinder wall 
 at  infinity $\mathfrak{Z}_\infty$.  The flux is confined in the (multiple) vortices
which are infinitely far away from the cylinder wall $\mathfrak{Z}_\infty$ and therefore vanishes
exponentially with correlation length proportional to $gv$, see (\ref{BPS}). Hence at the 
cylinder wall one has asymptotic vortex behavior, with winding in the squarks and the 
long-ranged gauge fields. Due to the monopoles this winding depends on $x^3$, but this dependence
is exponentially located at the monopoles with the characteristic length given by the associated mass
difference $|\Delta m|$. Concretely, the asymptotic field behavior is as follows:
\\

\noindent
{\underline{Cylinder Wall $\mathfrak{Z}_\infty$:}}  The confinement of the monopoles/flux by vortices implies   
$\ca{B}_k \approx 0 \approx \ca{A}_r$, where $\approx$ means equal up to
exponentially suppressed terms $\sim e^{-gvr}$. The Higgs fields, i.e. the squarks $\Sigma_i$ 
approach their vacuum values (up to winding) exponentially fast, i.e. with the spatial angular 
coordinate $\theta$ one has,
\begin{equation}
  \label{asym1}
  \Sigma_i\approx U q_i^{\mathrm{vac}} \quad , \quad U = e^{i\theta (w+w_{C+F})} \,  ,
\end{equation}
where the part of $U$ lying in the unbroken color-flavor symmetry $H_{C+F}$ (\ref{HCF}), generated
by $w_{C+F}(x^3)$, has a kink-like localized $x^3$-dependence at the monopoles so that $U$ is in the 
Cartan subgroup for $x^3\rightarrow \pm\infty$, see \cite{Shifman:2004dr}
for an explicit example. 
The asymptotic form of the BPS equations (\ref{BPS}) determines the residual fields as
\begin{equation}
  \label{asym2}
  \ca{A}_3 \approx i U\del_3\, U^{-1}\ \ ,\ \ \ca{A}_{\theta}\approx i U\del_\theta\, U^{-1}\ \ , \ \ 
  \vphi \approx U\phi_0U^{-1}\ ,
\end{equation}
where we note also that in addition to the  BPS equations (\ref{BPS}) one consequently 
has $\ca{D}_{\bar z}\Sigma_i \approx \ca{D}_{\bar w}\Sigma_i \approx \aux{D}\approx 0$ asymptotically. 
\\

\noindent
{\underline{Discs $\mathfrak{D}_\pm$:}} At the discs at infinity the fields approach 
 pure vortex behavior exponentially fast, with suppressed corrections $\sim e^{-|\Delta m \Delta x^3|}$, 
$\Delta x^3$ being the distance to the monopoles which goes to infinity at $\mathfrak{D}_\pm$. 
Therefore one has $\ca{B}_1\approx\ca{B}_2\approx\ca{A}_3\approx\ca{A}_r\approx 0$ and $\vphi\approx \phi_0$. 
The nontrivial fields $\ca{B}_3, \ca{A}_\theta$ in general take different values at the two  discs 
at infinity, but not the abelian $U(1)$ part. In particular one has 
\begin{equation}
  \label{asym3}
  \Sigma_i{}^m|_{\mathfrak{D}_\pm} \approx 
  \mathrm{diag} (\sigma_1^{\pm},\ldots, \sigma_N^{\pm})\ \ , \ \ 
   \ca{A}_{\theta}|_{\mathfrak{D}_\pm} = \ca{A}^{(\pm)}_{\theta} \ \ , \ \ 
   \mathrm{Tr}\, \ca{A}_{\theta}|_{\mathfrak{D}_{+}} =  \mathrm{Tr}\, \ca{A}_{\theta}|_{\mathfrak{D}_{-}}\, ,
\end{equation}
and analogously for $\ca{B}_3|_{\mathfrak{D}_{\pm}}$. The BPS equations (\ref{BPS}) imply then 
$[\ca{A}_{\alpha=1,2},\, \phi_0] \approx 0$ and one has in addition 
$\ca{D}_{\bar w} \Sigma_i \approx 0$ since $(\vphi +m_i)\Sigma_i\approx 0$, where there is
no summation over the flavor index here. 
\\

\noindent
{\bf Classical Energy.} The classical energy density was given in (\ref{hstat}) where 
for the BPS background only the last three terms, the local charges (\ref{zbkg}), are non-vanishing. 
It is easy to see that with the asymptotic behavior just described the integrated domain wall charge 
$\ca{T}^{\mathrm{w}\,3}_3$ vanishes for any volume with compact extent in the $x^1, x^2$ direction. 
The same applies to the second term in the integrated vortex charge  $\ca{T}^{\mathrm{v}\,3}_3$, which
vanishes for any volume which is compact in the $x^3$-direction. The remaining terms give 
for the classical energy
\begin{equation}
  \label{Ecl}
  E_{\mathrm{cl}} = \int d^3x\ \ca{H}^{stat} = 2\pi v^2 \, k^{\mathrm{vor}} L 
     + \fr{2}{g^2}\int d^2x\ \mathrm{Tr}\, \{\sqrt{2}\ \phi_0\, \ca{B}_3\, \} \big{|}^{x^3=\infty}_{x^3=-\infty}\ ,
\end{equation}
where the first term is the total vortex tension times the regulated extent of the $x^3$ direction. The
total vortex number $k^{\mathrm{vor}} = \frac{1}{2\pi}\int d^2x \mathrm{Tr}\,\ca{B}_3 =\mathrm{Tr}\, w$, 
see (\ref{asym2}), was introduced previously and is positive according to our choices
for the signs in the BPS equations. 

The second term gives the magnetic charge and thus the mass of the confined monopole. It is determined
by the difference in the flux through the discs $\mathfrak{D}_{\pm}$. Following \cite{Nitta:2010nd} one
can express this flux in terms of the individual contributions to the vortex number according to the symmetry breaking pattern 
(\ref{vac4}). Corresponding to the group $H_{C+F}$ (\ref{HCF}) there are $q$ distinct topological 
quantum numbers
\begin{equation}
  \label{ktop}
  k_r^{\mathrm{vor}} = \frac{1}{2\pi}\int d^2x \ \mathrm{Tr}\,\{\ca{B}_3\, t_r^0\,\}
    \quad\Rightarrow\quad \sum_{r=1}^q  k_r^{\mathrm{vor}} =  k^{\mathrm{vor}} \ ,
\end{equation}
where $t_r^0= \mathrm{diag}(0,\ldots, \unit_{n_r},\ldots0)$ is the $U(1)$ generator of the $r$'th 
factor of  $H_{C+F}$ (\ref{HCF}). The vev of the adjoint scalar 
(\ref{vac3}) can thus be written as $\phi_0 = - \sum_{r=1}^q m_r\, t_r^0$.  
The monopole mass contribution to the classical energy 
is then given by
\begin{equation}
  \label{Mcl}
  M^{\mathrm{mon}}_{cl}= -\frac{4\pi}{g^2} \sum_{r=1}^q \sqrt{2}\, m_r \Delta\,k_r^{\mathrm{vor}}
     \quad\quad\textrm{with}\quad\quad
    \Delta\,k_r^{\mathrm{vor}}:= k_r^{\mathrm{vor}}|_{\mathfrak{D}_{+}}  -  k_r^{\mathrm{vor}}|_ {\mathfrak{D}_{-}} \,,
\end{equation}
which implies $\sum_{r=1}^q \Delta k_r^{\mathrm{vor}} = 0$.

For example, the approximate solution for $N=2$ given in \cite{Shifman:2004dr} obeys 
$\ca{A}_\theta|_{\mathfrak{D}_{\pm}} = \frac{1}{2}(\unit \pm \tau^3 )$ so that 
$\Delta\,k_1^{\mathrm{vor}}= -  \Delta\,k_2^{\mathrm{vor}} =1$.
The monopole mass is then given by $M^{\mathrm{mon}}_{cl}= \frac{4\pi}{g^2} \sqrt{2}\,(m_2-m_1)$.

\subsection*{Quantum Correction}

To compute the one-loop energy correction for the confined monopole one needs the $T_{00}$ 
component of the energy momentum tensor in second order of the quantum fields. For 
the sake of regularization we consider the six-dimensional expression (\ref{TMN}), keeping the 
$x^5$ dependence of the fields as we did in the previous section. 

It turns out to be convenient to add a BRS-exact piece to the energy momentum tensor, namely the
contribution from the gauge fixing and the ghosts (\ref{gf}). One finds similar simplifications to those found
for the Lagrangian. As mentioned above, (\ref{TMN}) describes the gravitational energy momentum 
tensor. To add the appropriate gauge-fixing and ghost tensor we compute the gravitational effect
of the Lagrangian (\ref{gf}). To do so we embed  it in curved space and vary w.r.t.\ the metric,
\begin{align}
  \label{Tgf}
  T_{MN}^{\mathrm{gf+gh}} = -2\, \frac{\delta S^{\mathrm{gf+gh}}}{\delta g^{MN}}\big{|}_{\eta_{MN}}
   = -2\, \delta_{BRS}\, \frac{\delta}{\delta g^{MN}} \int d^Dx\,\sqrt{-g}\ 
      \Upsilon_{\mathrm{grav}}\big{|}_{\eta_{MN}}\ ,
\end{align}
where we have used that the BRS transformation obviously commutes with the variation w.r.t.\ the metric, which
is evaluated for the flat metric $\eta_{MN}$. The overall factor matches the convention for 
(\ref{TMN}). The gauge-fixing fermion $\Upsilon$ (\ref{gf}) in curved space is (we set $\zeta=1$),
\begin{equation}
  \label{gffgr}
  \Upsilon_{\mathrm{grav}}=  \fr{2}{g^2}\mathrm{Tr}\, \{\, b \big{[}\, \fr{1}{\sqrt{-g}} 
    \ca{D}_M (\sqrt{-g}\, g^{MN}a_M )
   +\fr{i\, g^2}{2} \,( S_{iI}\otimes \bar{\Sigma}^I_i - 
     \Sigma_{iI}\otimes \bar S_i^I) +\fr{1}{2} B \big{]}\},
\end{equation}
where $a_M= A_M-\ca{A}_M$, with $\ca{A}_M$ and $\Sigma_{iI}$ some functions. Variation w.r.t.\ 
the metric and BRS transformation gives for the 
gauge-fixing energy momentum tensor
\begin{align}
  \label{Tgf2}
    T_{MN}^{\mathrm{gf+gh}} =&\  \eta_{MN}\big{(}\,\ca{L}^{\mathrm{gf+gh}} 
     - \fr{2}{g^2}\,\delta_{BRS}\, \del_P \mathrm{Tr}\, \{\, b\, a^P\}\big{)}\nonumber\\[3pt]
     &\hspace{-0.15cm}-  \fr{4}{g^2}\mathrm{Tr}\, \{ \ca{D}_{(M}\, b D_{N)}\,c 
    + a_{(M}\ca{D}_{N)}[\, \ca{D}_{P}a^{P} +\fr{i\, g^2}{2} \,( S_{iI}\otimes \bar{\Sigma}^I_i - 
     \Sigma_{iI}\otimes \bar S_i^I)  \,] \} \,
\end{align}
where the first line contains the usual Lagrangian contribution and a BRS-exact total derivative term. In 
choosing the background functions $\ca{A}_M, \Sigma_{iI}$ to be the BPS-background (\ref{BPS})
this total derivative gives exactly the first line in the total derivative terms for the quadratic
Lagrangian (\ref{totd}), which can therefore in general be safely omitted.

Expanding the total energy density $T_{00}^{\mathrm{tot}}= T_{00}+T_{00}^{\mathrm{gf+gh}}$, to second order 
in the quantum fields around the BPS background (\ref{BPS}) and keeping the dependence on the regulating
extra dimension $x^5$ one obtains
\begin{align}
  \label{T2}
  T_{00}^{\mathrm{tot}} =&\ \ca{H}^{\mathrm{stat}}_{cl} 
                + 2\, \mathrm{Tr}\, \{\,\del_0 W^\dagger\del_0 W\, \} 
       + \fr{i}{2}\, \mathrm{Tr}\, \{\,\Psi^\dagger\del_0\Psi-\del_0\Psi^\dagger\Psi\,\} \nonumber\\[3pt]
      &\hspace{1.1cm} +\fr{2}{g^2} \mathrm{Tr}\, 
       \{ (\del_0 a_5)^2 + (a_0\del_0^2a_0) -2(\del_0b\,\del_0c)\}\nonumber\\[3pt]
     & +\fr{1}{g^2}\, \del_k^2\, \mathrm{Tr}\, \{a_0^2- bc\,\} +\del_k \mathrm{Tr}\, \{W^\dagger\ca{D}_k W\,\}
     -\fr{1}{g^2}\, \del_k\del_l \mathrm{Tr}\, \{a^k a^l\,\} +O(3)\, ,
\end{align}
where we omitted total derivative terms of fields which are not connected by the given propagators 
and we used the previously derived fluctuation equations for the Lagrangian part. 
The first line gives the bulk contribution, the fields $W$ and $\Psi=(U,V)$ were defined above in (\ref{W}), 
(\ref{UV}). The contributions of the quartet in the second line obviously cancel each other upon taking the expectation value.
This is analogous to the situation for the monopole in the Coulomb phase \cite{Rebhan:2006fg}. The last line 
collects all total derivative terms, $O(3)$ stands for higher orders in the quantum fields.
Here we will focus on the bulk contributions. As already mentioned, the total derivative terms may contribute 
to composite operator renormalization. This does not happen in the Coulomb phase for $\sN=2$ (but it does for $\sN=4$),
see \cite{Rebhan:2006fg}. However, as discussed for the classical solution above, the geometry of the boundary at 
infinity and the asymptotic behavior of the fields is rather particular and different from the Coulomb phase
and these surface terms deserve a detailed analysis which we leave for a future investigation. 
\\

\noindent
{\bf Bulk Quantum Correction.} For the bulk contribution to the expectation 
value of $ T_{00}^{\mathrm{tot}}$ in the BPS sector,  one can either directly insert the 
mode expansions for the fields or use the propagators defined above (\ref{propW}) and (\ref{propF})
in the form\footnote{To carry out the $p^0$-integration in the propagators one already has to use 
the formulas from dimensional regularization, see for example \cite{Peskin:1995ev}. 
In fact this is just a formality 
that is necessary only because we chose, for compactness of the representation, to use (off-shell)
propagators. Inserting directly the (on-shell) mode expansion of the quantum fields no such issue
arises.} 
\begin{equation}
  \label{expW}
  \langle \, \mathrm{Tr}\, \{\,\del_0 W^\dagger\del_0 W\,\}\,\rangle = 
   \lim_{x'\rightarrow x}  \mathrm{Tr}\, \{\,\del_0\,\del_0'\, \langle \,W(x) \,W^\dagger(x')\,\rangle\,\}\, ,
\end{equation}
and similarly for the fermions, where the trace on the r.h.s.\ includes the trace over the component
indices of $W$. The resulting one-loop correction is obtained as,
\begin{align}
  \label{Ebulk}
  \Delta E^{(1)}_{\mathrm{bulk}} =&\ \int d^3x\, \langle\, T_{00}^{\mathrm{bulk}}\,\rangle \nonumber\\
     =&\ \frac{1}{2} \int\frac{d^{\epsilon}\ell}{(2\pi)^\epsilon} \int_n\!\!\!\!\!\!\!{\textstyle{\sum}}\, 
    \mu_n\,\ \sqrt{\omega_n^2+ \ell^2} \int d^3 x \ 
    ({\lVert v_n\rVert}^2 - {\lVert u_n\rVert}^2 )\nonumber\\
 =&\ \frac{1}{2} \int\frac{d^{\epsilon}\ell}{(2\pi)^\epsilon} 
    \int_n\!\!\!\!\!\!\!{\textstyle{\sum}}^{\mathrm{cont}}\, 
    \mu_n\,\ \sqrt{\omega_n^2+ \ell^2} \  \Delta\rho_n \, ,
\end{align}
where $\lVert v_n\rVert^2 = \mathrm{Tr}\,\{ v^\dagger_n(x)\, v_n(x)\}$, etc.
In the second line we used that the contribution from the zero-modes vanish separately, 
since these are  in fact massless modes with momenta $\ell$ and give scaleless integrals 
that do not contribute in dimensional regularization. In the last line
we used that the discrete non-zero modes come in $u,v$-pairs with well defined normalization 
$\int d^3x\,\lVert v_n\rVert^2 = \int d^3x\,\lVert u_n\rVert^2 =1$ and therefore 
only the continuum modes remain, as indicated. 
For the continuum modes, where the spatial integral is defined only for the difference  
of $u, v$ modes, we introduced the spectral density (difference)
\begin{equation}
  \label{specd}
  \Delta\rho_{\{n_s\}} :=  \int d^3 x \  ({\lVert v_{\{n_s\}}\rVert}^2 - {\lVert u_{\{n_s\}}\rVert}^2 )\ ,
\end{equation}
where we emphasized the dependence on all discrete and continuum quantum numbers $\{n_s\}$.
This spectral density is the key quantity for computing one-loop corrections in solitonic backgrounds.
Note that we did not introduce any compact volume here, it will turn out that in contrast to the 
classical energy the quantum correction is well defined in this case.
\\

\noindent
{\underline{Spectral Density, Index Theorem:}} Index theorems have turned out to be important not only
for determining the classical moduli for monopoles and such, but have also proven to be powerful
in quantum calculations. The reason for this is that spectral densities such as (\ref{specd})
can be extracted from them. 

The index of an operator $L\,$, $\mathrm{Ind}(L) = n_{L^\dagger L}^0 - n_{L L^\dagger}^0\,$, 
counts the difference in the number of zero modes $n^0$ of $L$, and $L^\dagger$, and can be obtained
from an IR-regulated expression:
\begin{equation}
  \label{ind1}
  \ca{I}(M^2):= \mathrm{T}_{\mathrm{R}}\,\Big{\{} \frac{M^2}{L^\dagger L+M^2}-\frac{M^2}{LL^\dagger+M^2}\,\Big{\}}
   \ \ \  \Rightarrow \ \ \ 
   \mathrm{Ind}(L) = \lim_{M^2\rightarrow 0} \ca{I}(M^2)\ ,
\end{equation}
where the modified symbol for the trace indicates that the trace is taken now also over the functional 
Hilbert space. Of particular interest for us here is the application in non-compact spaces 
as developed in \cite{Callias:1977kg,Weinberg:1979ma,Weinberg:1981eu}. The contribution from the 
continuum states to the quantity 
$ \ca{I}(M^2)$ is related to the sought after spectral density in the following way,
\begin{equation}
  \label{ind2}
  \ca{I}^{\mathrm{cont}}(M^2) := \ca{I}(M^2)-\ca{I}(0) = 
   \int_n\!\!\!\!\!\!\!{\textstyle{\sum}}^{\mathrm{cont}}\, \mu_n\, \frac{- M^2}{\omega_n^2+M^2}\ \Delta \rho_n\, ,
\end{equation}
where the minus is introduced to match the definition for $\Delta\rho_n$ given in (\ref{specd}) and
we introduced the same measure factor $\mu_n$ which will come out from the explicit computation.
Denoting the eigenmodes of $LL^\dagger, L^\dagger L$ by $v_n,u_n$ it is easy to see that this definition 
of the spectral density coincides with the one given in (\ref{specd}).

The index has been computed
for numerous different topological solutions in different models, in the given context  
for example for vortices and domain walls \cite{Hanany:2003hp,Sakai:2005sp}, but not 
for confined monopoles. 
The usual technique, in the case of nontrivial $M$-dependence,  
is to transform $\ca{I}(M^2)$ into a surface term which is therefore
determined by the asymptotic values of the background fields,\footnote{Using the susy quantum 
mechanical relations  (\ref{fmod3}) one can express directly in (\ref{specd}) 
the $u$-modes through the $v$-modes, or vice versa,  plus a surface term that depends on the modes, 
see for example \cite{Rebhan:2003bu}. The advantage of the index theorem is that the resulting surface term is 
completely determined by the asymptotic background alone and no knowledge, not even about the asymptotic  
behavior, of the modes is necessary.} i.e. their topological properties. In the considered cases one 
of the two operators $L,L^\dagger$
is strictly positive, i.e. has no zero modes. 
In the following $L,L^\dagger$ stand of course for our operators (\ref{LLd}) but their 
properties are different from the usual situation.
Firstly, neither of the operators (\ref{LLd}) is manifestly strictly positive \cite{Wimmer:2011}. The 
second difference is due to the axial geometry of confined monopoles and the nontrivial behavior
at the cylinder at infinity, see the discussion at the beginning of this section. In particular at the 
discs at infinity $\mathfrak{D}_{\pm}$ the background has the full nontrivial form of (multiple) 
vortices. In a sense one needs a second ``index theorem'' for the resulting boundary term on these 
discs. The necessary generalization of index theorem calculations and techniques are developed and
applied to the given case in \cite{Wimmer:2011}. We will only take the result of this analysis which 
is needed for the 
considerations here.

For a general background with the topological properties as given in (\ref{ktop}),  
(\ref{Mcl}) the result for the (regulated) index (\ref{ind1}) is
\begin{equation}
  \label{ind3}
  \ca{I}(M^2)= -\fr{1}{2}\, \sum_{r=1}^q\ \Delta k_r^{\mathrm{vor}}\, \Big{(}\, 2\sum_{i=1}^N - \sum_{i=1}^{\Nf}\, \Big{)}
    \frac{\sqrt{2} (m_i-m_r)}{\sqrt{2(m_i-m_r)^2+M^2}}\ .
\end{equation}
For further details and discussion of this index we refer to \cite{Wimmer:2011}. This gives an obvious
identification for summation over the discrete quantum numbers in (\ref{Ebulk}) and the dependence 
of the mode energies $\omega_n$ on them. The dependence on the continuum quantum numbers, i.e. momenta,
is obtained according to (\ref{ind2}):\footnote{The assumed axial geometry of the classical background
implies of course a dependence solely on the radial momentum. In deference to
the conventions used for the Coulomb monopole \cite{Rebhan:2004vn} we have rewritten the momentum integration
in three-dimensional space. Furthermore, the contribution to the index $\ca{I}(0)$ vanishes for $i>N$, 
nevertheless it is necessary to formally keep this vanishing contribution to obtain a $M$-independent 
spectral density.}
\begin{align}
  \label{ind4}
   \int_n\!\!\!\!\!\!\!{\textstyle{\sum}}^{\mathrm{cont}}\, \mu_n\, \Delta\rho_n\ \rightarrow \ 
           \sum_{r=1}^q\ \Delta k_r^{\mathrm{vor}}\, \Big{(}\, 2\sum_{i=1}^N - \sum_{i=1}^{\Nf}\, \Big{)}
          \int\frac{d^3k}{(2\pi)^3} \ \frac{-2\pi\sqrt{2}(m_i-m_r)}{k^2[k^2+2(m_i-m_r)^2]}
\end{align}
with the mode energies $\omega^2_n = k^2 + 2(m_i - m_r)^2$. Inserting these results one obtains for 
the energy correction (\ref{Ebulk}),
\begin{align}
  \label{delE}
  \Delta E_{\mathrm{bulk}}^{(1)} =&\ -\pi \sum_{r=1}^q\  \Big{(}\, 2\sum_{i=1}^N - \sum_{i=1}^{\Nf}\, \Big{)}
       \Delta k_r^{\mathrm{vor}}\, \sqrt{2}\,(m_i - m_r)  
      \nonumber\\ 
    &\hspace{2cm} \times \int\frac{d^\epsilon\ell}{(2\pi)^\epsilon}\frac{d^3k}{(2\pi)^3} 
    \frac{\sqrt{k^2+\ell^2+2(m_i-m_r)^2}}{k^2[k^2+2(m_i-m_r)^2]}\ .
\end{align}
The integral in the second line has to be carried out in dimensional regularization. The resulting expression
is
\begin{align}
  \label{dimreg}
    \frac{1}{4\pi^2}(1-\epsilon)&\left[\frac{1}{(4\pi)^{\epsilon/2}}\,\Gamma\big{(}-\frac{\epsilon}{2}\big{)}
            [\,2(m_i-m_r)^2\,]^{\epsilon/2}\right] = \nonumber\\[3pt]
            &\hspace{2.5cm} \frac{1}{2\pi^2} 
           + \frac{1}{4\pi^2}\log \left[\frac{M_{UV}^2}{\mu^2}\right]
           - \frac{1}{4\pi^2}\log \left[ \frac{2(m_i-m_r)^2}{\mu^2}\right]\, ,
\end{align}
where we have introduced the renormalization scale $\mu$ which is of the order of the masses
$m_i$ and we employed the $\overline{MS}$-scheme 
$\log(M_{UV}^2/\mu^2)=-\frac{2}{\epsilon}-\gamma-\log(\mu^2/4\pi)$, see e.g. \cite{Peskin:1995ev}, 
but as discussed 
above, our regularization is effectively the dimensional reduction scheme $\overline{DR}$. 
The first term on the r.h.s.\
in (\ref{dimreg}) is a renormalization scale invariant constant which stems from the explicit factor 
$\epsilon$ on the l.h.s.\ and will be identified with an anomaly. 

Inserting (\ref{dimreg}) in (\ref{delE}) the total one-loop (bulk) energy including the classical
energy (\ref{Ecl}), (\ref{Mcl}) is obtained as,
\begin{align}
  \label{E1}
  E^{\mathrm{pert}} =&\
      2\pi v^2 k^{\mathrm{vor}}L  - \frac{4\pi}{g^2{\,(\mu)}} 
         \sum_{r=1}^q \sqrt{2}\,m_r \Delta k_r^{\mathrm{vor}}
     + \frac{2N-N_f}{2\pi}  \sum_{r=1}^q \sqrt{2}\,m_r \Delta k_r^{\mathrm{vor}} \nonumber\\[3pt]
     &+ \frac{1}{4\pi} \sum_{r=1}^q\ \Delta k_r^{\mathrm{vor}}\, \Big{(}\, 2\sum_{i=1}^N - \sum_{i=1}^{\Nf}\, \Big{)} 
     \sqrt{2}\,(m_i-m_r)\log \left[ \frac{2(m_i-m_r)^2}{\mu^2}\right]\, ,
\end{align}
where we have indicated that this result includes the complete perturbative quantum correction which is
one-loop exact due to $\sN=2$ supersymmetry. We also 
introduced the renormalized coupling constant for $\sN=2$ SQCD, see e.g. \cite{Peskin:1997qi},
\begin{equation}
  \label{reng}
\frac{4\pi}{g^2{\,(\mu)}} = 
  \frac{4\pi}{g^2{\,({\scriptstyle{M_{UV}}})}} - 
       \frac{2N-N_f}{4\pi}\log \left[\frac{M_{UV}^2}{\mu^2}\right] 
       = \frac{2N-\Nf}{2\pi}\log\left[\,\frac{\mu}{\Lambda}\,\right]\, ,
\end{equation}
where in the last equality we traded the coupling constant for the dynamically generated 
RG-group invariant scale $\Lambda$. See \cite{Finnell:1995dr} for how this scale is related to other 
regularization/renormalization schemes. 
We however emphasize that for a consistent computation of quantum corrections in the soliton 
background, especially to obtain the renormalization scale invariant anomaly in (\ref{E1}), 
the regularization method employed here is crucial. 

The first two terms in (\ref{E1}) are just the classical vortex tension and monopole mass, 
(\ref{Ecl}), (\ref{Mcl}). The third term, which is proportional to the $\beta$-function coefficient
$b_0= 2N-\Nf$,   
is an anomalous contribution, whereas the last line gives renormalization scale dependent corrections. 
The perturbative corrections do not need any IR-regularization by a compact volume like the classical vortex 
tension does, in fact there are no volume proportional corrections. Thus the vortex tension of the 
confined monopole is 
not affected by (perturbative) quantum corrections which is also in agreement with the fact that the 
FI-term does not renormalize in $\sN=2$ SQCD \cite{Argyres:1996eh}.  

As an example, we evaluate the quantum-corrections in (\ref{E1}) for a single confined monopole for 
$N=\Nf=2$. The classical mass for this case was given below (\ref{Mcl}) and we set 
$\sqrt{2}\,(m_2-m_1) = m$. Here we note that the anomalous contribution in (\ref{E1}) can be written as 
$M_{\mathrm{ano}}^{\mathrm{mon}}= -\frac{g^2(\mu)}{8\pi^2}\,b_0\,M_{cl}^{\mathrm{mon}}$.
Choosing the renormalization scale equal to the single mass parameter, i.e. 
$\mu = m$, the energy of a single confined monopole is then
\begin{equation}
  \label{mon}
   E^{\mathrm{pert}} =\
      2\pi v^2 k^{\mathrm{vor}}L  + \frac{4\pi\, m}{g^2{\,(\mu)}} - b_0\,\frac{m}{2\pi}\ ,
\end{equation}
with $b_0 = N = 2$ for the considered example, in which case the correction equals the quantum correction 
for the $CP^1$ kink \cite{Shifman:2006bs, Mayrhofer:2007ms}. Extending this result to the case of pure $\sN=2$ 
Yang-Mills theory, i.e. setting $\Nf=0$ and thus $b_0=4$, 
the correction exactly matches that obtained for the Coulomb monopole in 
$\sN=2$ SYM, which is given solely by the anomaly term
and was first obtained in \cite{Rebhan:2004vn}. 
The renormalization 
scale chosen here coincides with the renormalization condition of
\cite{Rebhan:2004vn,Rebhan:2006fg}.\\

\noindent
{\bf{Central Charge Anomaly.}} We briefly mention the relation of the discussed correction
to an anomaly in the central charge. In \cite{Rebhan:2002yw} it was shown for the two-dimensional kink 
that the central charge gets an anomalous contribution from the \emph{canonical} 
fermionic momentum operator in the 
regulating extra dimension. In complete analogy an anomalous contribution was discovered in 
the central charge for the Coulomb monopole and it was shown that this contribution is 
necessary for BPS saturation and consistency with the Seiberg-Witten low energy effective 
action \cite{Rebhan:2004vn}.

As explained below (\ref{ac}) the fermionic $U$-  and $V$-zero modes are in fact massless 
modes with momentum in the regulating extra dimensions. However, only particles
with momentum in a fixed direction can be created, where massless $U$-particles 
propagate in the opposite direction to massless $V$-particles. The non-vanishing of 
the index (\ref{ind3}), which counts the difference in the $U$- and $V$-zero modes, shows that
there is a mismatch which leads to a non-vanishing current in the extra dimension
with a finite remainder when the regulator is removed. This explains the occurrence  
of the anomaly in the central charge and its location in the fermionic momentum operator
in the regulating extra dimension.

The central charge (density) operator  (\ref{Z}) which carries the anomaly is obtained 
by dimensional reduction of the six-dimensional susy algebra (\ref{dJ}). The explicit six-dimensional
origin reads
\begin{equation}
  \label{za1}
   \ca{Z} = \frac{1}{\sqrt{2}} \big{[} (T^{05}+\vep^{0ijk}\ca{Z}_{ijk6} )
                  - i(T^{06} - \vep^{0ijk}\ca{Z}_{ijk5}) \big{]} \, ,
\end{equation}
where in the trivial dimensional reduction derivatives w.r.t.\ the regulating $x^5$
direction are neglected (all fields are independent of $x^6$ even when regulated). The expected
anomaly from a fermionic current in the extra-dimension is therefore given by 
the canonical momentum operator in the regulating 
extra-dimension, which is proportional to $\del_5$:
\begin{equation}
  \label{za2}
  \sqrt{2}\, \langle Z\rangle_{\mathrm{ano}} =\int_{vol} \langle T^{05\ \mathrm {can}}_{\mathrm{ferm}}\rangle\, .
\end{equation}
The canonical fermionic momentum operator is obtained from the gravitational stress tensor 
(\ref{TMN}) by replacing the explicit symmetrization (which is of weight one) by a factor two,
the difference is an antisymmetric tensor.\footnote{
An explicit calculation shows that the symmetric energy momentum operator (\ref{TMN})
gives the anomaly plus further finite renormalization scale dependent contribution of 
the same form as for the energy (\ref{E1}). That is, the mentioned antisymmetric difference between 
the canonical and the symmetric energy tensor gives ordinary corrections to the central charge that vanish 
for $N=N_f=2$ as  in (\ref{mon}).}
Using condition (\ref{6dspin}) to express $\lambda_1$ in terms of $\lambda_2$ and the 
definition of the $U$ and $V$ fields (\ref{UV}), (\ref{fqf}) the relevant operator 
reads $ T^{05\ \mathrm {can}}_{\mathrm{ferm}} = -\frac{i}{2}\, \mathrm{Tr}\, \{\,
\Psi^\dagger \overset{\leftrightarrow}{\del}_5 \Psi\,\}$. Analogous to (\ref{expW})
we use the fermionic propagator (\ref{propF}) to obtain
\begin{align}
  \label{za3}
  \int_{vol} \langle T^{05\ \mathrm {can}}_{\mathrm{ferm}}\rangle\ &= 
    \frac{1}{2} \int\frac{d^{\epsilon}\ell}{(2\pi)^\epsilon} 
    \int_n\!\!\!\!\!\!\!{\textstyle{\sum}}^{\mathrm{cont}}\, 
    \mu_n\,\ \frac{\ell^2}{\sqrt{\omega_n^2+ \ell^2}} \  \Delta\rho_n \nonumber\\[7pt]
   &= -\frac{1}{2}  \sum_{r=1}^q\ \Delta k_r^{\mathrm{vor}}\, \Big{(}\, 2\sum_{i=1}^N - \sum_{i=1}^{\Nf}\, \Big{)}
     \sqrt{2}(m_i-m_r)\times\nonumber\\[5pt]
    &\hspace{8mm}
      \int\frac{d^\epsilon\ell}{(2\pi)^\epsilon}\frac{dk}{2\pi} 
    \frac {\ell^2}{[k^2+2(m_i-m_r)^2]\sqrt{k^2+\ell^2+2(m_i-m_r)^2}}\,.
\end{align}
In the second line we inserted the spectral density and mode energies according to (\ref{ind4}).
The integral, to be evaluated in dimensional regularization, is of exactly the same form
as for the central charge anomaly for the Coulomb monopole in $\sN=2$ SYM and its value 
is independent of the masses and given by $\frac{1}{\pi}$  \cite{Rebhan:2004vn}. The 
central charge anomaly is then given by
\begin{equation}
  \label{za4}
  \langle Z\rangle_{\mathrm{ano}} =\frac{1}{ \sqrt{2}}\ \frac{2N-N_f}{2\pi}  
       \sum_{r=1}^q \sqrt{2}\,m_r \Delta k_r^{\mathrm{vor}} = -\,b_0\,\frac{g^2(\mu)}{8\pi^2}\ Z_{cl}\, ,
\end{equation}
where we have again used that $ \sum_{r=1}^q\Delta k_r^{\mathrm{vor}} = 0$. The last expression is 
of the form which is generally valid, also for the Coulomb monopole. Comparing 
with the second term in the energy (\ref{E1}) one sees that the anomalies separately satisfy the 
BPS condition (\ref{bps4}), i.e. $\sqrt{2}\,\langle Z\rangle_{\mathrm{ano}} = M_{\mathrm{ano}}^{\mathrm{mon}}$ .  

This is of course a rather incomplete discussion of the central charge corrections, it is 
a particular feature of our regularization method that the anomaly is the part of the correction
which is obtained in the simplest way.  
A detailed analysis of the central charge correction, in particular the 
boundary terms mentioned above, as well as the anomaly multiplet structure will be given elsewhere. 

\subsection{Comparison with  $CP^{n}$ Models}  

From the early days of QCD on it was observed that $CP^{n}$ models capture many 
aspects of QCD 
(confinement, generation of a mass gap, asymptotic freedom \ldots) and were used
as a laboratory for the study of strongly coupled QCD. In \cite{Dorey:1998yh} 
it was observed that for $\sN=2$ SQCD (with $N_f=N$) this relation holds even at a quantitative
level. There is an exact match between the BPS spectra of $\sN=2$ SQCD on the Coulomb
branch (known from SW solutions) and $\sN = (2,2)$ $CP^{N-1}$ models in two dimensions.
Subsequently this was explained by the fact that the effective low-energy
theory of non-abelian vortices is given by an $\sN = (2,2)$ $CP^{N-1}$ model
on the vortex world sheet \cite{Hanany:2004ea,Shifman:2004dr}. Though the vortices
exist only in the Higgs phase the vev for the adjoint scalar (\ref{vac3}) is exactly
the root of the baryonic Higgs branch \cite{Argyres:1996eh} and thus continuously connected,
in parameter space (via the FI-parameter), to the Coulomb branch. However, there is of course 
a nontrivial transition involved which confines the monopoles and is for example reflected
in the volume proportional energy contribution in (\ref{E1}) from the confining vortices.

Here we compare the BPS spectrum of kinks in the $CP^{N-1}$ model (with twisted masses) 
with our results for confined monopoles which are obtained directly in the Higgs phase. 
The superpotential for the $\sN = (2,2)$ $CP^{N-1}$ model is known exactly 
due to holomorphicity \cite{Hanany:1997vm, Dorey:1998yh} (see also \cite{Tong:2008qd}\footnote{There is an 
obvious factor $\frac{1}{2}$ missing in \cite{Tong:2008qd} which is clear by comparing the
 classical energies ibidem.}). Including non-degenerate twisted masses it reads:
\begin{equation}
  \label{cp1}
  \ca{W}(\sigma)= - r(\mu)\,\frac{\sigma}{2} -\frac{1}{4\pi}\sum_{i=1}^N\,(\sigma - m_i) 
 \left[\, \log\,\left(\frac{\sigma-m_i}{\mu}\right)-1\, \right]\, ,
\end{equation}
where $\sigma$ is a complex scalar and  $r(\mu)$ is related to the renormalized sigma model
coupling constant. More importantly,  $r(\mu)$ is related to the gauge coupling constant of the
four-dimensional theory as $r(\mu) = \frac{4\pi}{g^2(\mu)}$ \cite{Dorey:1998yh}. All our considerations
assume a vanishing vacuum theta-angle. The vacua are given by the critical points of $\ca{W}$ and are
thus determined by the condition
\begin{equation}
  \label{cp2}
  \prod_{i=1}^N(\sigma - m_i) = \Lambda^N \, .
\end{equation}
For the perturbative regime ($\Delta m_i\gg \Lambda$) the vacua are approximately given by their 
classical
values, which are the $N$ different points $\sigma_i=m_i$. The different vacua are connected by 
BPS kinks. The mass, and central charge, for a kink connecting the $r$'th with the $s$'th 
vacuum is then given by
\begin{align}
  \label{cp3}
  M_{rs}^{\mathrm{kink}} =&\ 2\,[\ca{W}(\sigma_r)-\ca{W}(\sigma_s)]\nonumber\\
   =&\ \frac{4\pi}{g^2(\mu)} (m_s-m_r)
     -\frac{N}{2\pi}(m_s-m_r) \nonumber\\ &\ \ \ + \frac{1}{4\pi}\sum_{i=1}^N
   \left[
      (m_i-m_r)\,\log\left[\frac{(m_i-m_r)^2}{\mu^2}\right] 
    - (r \rightarrow s)\right] \, ,
\end{align}
where we chose the branch cuts of the logarithm such that there are no electric charge contributions,
see \cite{Dorey:1998yh} for details. Multiplying the masses with the obvious factor 
$\sqrt{2}$, which comes from our convention for the super potential in (\ref{eq:lss}), and taking for the 
confined monopole the topological quantum numbers $\Delta k_r^{\mathrm{vor}}= - \Delta k_s^{\mathrm{vor}} = 1$ 
and otherwise zero the quantum energy for the confined monopole (\ref{E1}) gives for $N_f = N$:
\begin{equation}
  \label{cp4}
    E^{\mathrm{pert}}-2\pi v^2 k^{\mathrm{vor}}L =  M_{rs}^{\mathrm{kink}}\ .
\end{equation}
The confined monopole mass agrees with the kink mass. 

\section{Summary and Conclusions}

We studied several aspects of confined monopoles in $\sN=2$ SQCD with 
$SU(2)_R$ covariant FI-terms and gauge group $U(N)$. The main result of the 
paper is the formula for the quantum energies of multiple confined monopoles
(\ref{E1}), which is one-loop exact and thus contains the full perturbative information.
For the special case of $N_f=N$ we find complete agreement with the perturbative corrections
for kinks in $\sN=(2,2)$ $CP^{N-1}$ models. We also compute the anomalous contribution in the central
charge. However, the central charge sector needs a broader study which will be given elsewhere,
in particular the anomaly multiplet structure and possible composite operator renormalizations 
due to surface terms. Central to the computations presented here  is the regularization 
by embedding the theory in a six-dimensional model with $U(1)$ background fields to generate 
the masses and the generalized index theorem \cite{Wimmer:2011}. The study of the $SU(2)_R$ 
covariant representations including tensorial central charges 
and the associated $\frac{1}{4}$ BPS equations shows that the spatial orientation of the 
confined monopoles is completely independent of the $SU(2)_R$ orientation. Interestingly,
at the quantum level the spatial $SU(2)$ gets linked to the R-symmetry $SU(2)_R$. 
From a phenomenological point of view it is of greatest interest to extend the analysis
presented here 
to cases with less supersymmetry, in particular to the case of increasing value of the
adjoint mass parameter $\mu$.
\\


\noindent
{\bf Acknowledgments:}
We thank A. Rebhan and P. van Nieuwenhuizen for careful reading, useful comments
and collaboration in the early stages of the project and especially P. van 
Nieuwenhuizen for checking several equations. We also thank A. Ipp for
help with the graphics and R.W. thanks H. Samtleben for discussions. 
This work is supported in part by the Agence Nationale de la Recherche (ANR). 
D.B. was supported by FWF project no. P19958-N16.

\appendix

\section{Fierz Identities and Conventions}

We are dealing with fermions in six dimensions of differing chiralities. 
Parameterizing the chiralities of two spinors in the general form as 
$\Gamma_7\chi=x\,\chi$ and  $\Gamma_7\psi = y\, \psi$, with $x,y =\pm1$ 
one has the following Fierz identity for $x=y$ :
\begin{equation}
  \label{a1}
  \chi\otimes\bar\psi = - \frac{1}{8}\,( \bar\psi\Gamma^M\chi)\,\Gamma_M(\unit-x\Gamma_7)
          +\frac{1}{48}\,(\bar\psi \Gamma^{MNP} \chi)\, \Gamma_{MNP}\, .
\end{equation}
Whereas for the case of $y=-x$ one has
\begin{equation}
  \label{a2}
  \chi\otimes\bar\psi = \left[-\frac{1}{8}\, (\bar\psi\chi) 
   + \frac{1}{16}\, (\bar\psi\Gamma^{MN}\chi)\Gamma_{MN}\right](\unit+x\Gamma_7) \, .
\end{equation}

For two symplectic Majorana spinors $\psi_I, \chi_I$ (chiral or not chiral) the following identity 
holds:
\begin{equation}
  \label{a3}
  \bar\psi^I\Gamma^{A_1}\cdots\Gamma^{A_n}\chi_J = 
   (-)^n \big{[} \delta^I{}_J \bar\chi^K\Gamma^{A_n}\cdots\Gamma^{A_1}\psi_K - 
   \bar\chi^I\Gamma^{A_n}\cdots\Gamma^{A_1}\psi_J \big{]}\, ,
\end{equation}
in particular one has $\bar\psi^I\Gamma^{A_1}\cdots\Gamma^{A_n}\chi_I = (-)^n \bar\chi^I\Gamma^{A_n}\cdots\Gamma^{A_1}\psi_I$.

Our definitions of $\Gamma_7$ and also $\gamma_5$ and space-time $\vep$-symbols 
coincide with those of \cite{VanProeyen:1999ni} in both four and six dimensions.
Relations for gamma matrices used to derive the results in the main text can be found
there. For the decomposition of products of gamma matrices we used the Mathematica
package ``GAMMA'' \cite{Gran:2001yh}.


\bibliographystyle{JHEP}

\providecommand{\href}[2]{#2}\begingroup\raggedright\endgroup

\end{document}